\documentclass[useAMS,usenatbib]{mn2e}

\newcommand{\kepler}{{\it Kepler}}
\newcommand{\mearth}{{\it MEarth}}
\newcommand{\prot}{\ensuremath{P_{\rm rot}}}
\usepackage{graphicx}
\usepackage{amsmath}
\usepackage{color}
\usepackage{natbib}

\usepackage{comment}

\title[Rotation of \kepler\ field M-dwarfs]{Measuring the rotation period distribution of field M-dwarfs with \kepler}
\author[A. McQuillan, S. Aigrain and T. Mazeh]{A. McQuillan$^{1}$\thanks{E-mail:
amy.mcquillan@astro.ox.ac.uk}, S. Aigrain$^{1,2}$ and T. Mazeh$^{3,2}$\\
$^{1}$Department of Physics, University of Oxford, Oxford, OX1 3RH, UK\\
$^{2}$All Souls College, Oxford, OX1 4AL, UK\\
$^{3}$School of Physics and Astronomy, Raymond and Beverly Sackler, Faculty of Exact Sciences, Tel Aviv University, 69978, Tel Aviv, Israel}
\begin{document}

\date{Accepted 26 March 2013, Received 19 September 2012}

\pagerange{\pageref{firstpage}--\pageref{lastpage}} \pubyear{2013}

\maketitle

\label{firstpage}

\begin{abstract}  
 We have analysed 10 months of public data from the \kepler\ space mission
 to measure rotation periods of main-sequence stars with
 masses between 0.3 and 0.55\,$M_\odot$. To derive the rotational period
 we introduce the autocorrelation function
 and show that it is robust against phase and
 amplitude modulation and residual instrumental systematics.
 Of the 2483 stars examined, we detected rotation periods in 1570 (63.2\%),
 representing an increase of a factor $\sim$30 in the number of rotation period
 determination for field M-dwarfs. The periods range from 0.37--69.7 days, with amplitudes
 ranging from 1.0--140.8 mmags. The rotation period distribution is clearly bimodal, with
 peaks at $\sim 19$ and $\sim 33$ days, hinting at two distinct waves
of star formation, a hypothesis that is supported by the fact that 
slower rotators tend to have larger proper motions. The two peaks of the
rotation period distribution form two distinct sequences in
period-temperature space, with the period decreasing with increasing temperature,
reminiscent of the Vaughan-Preston gap.
The period-mass distribution of our sample shows no evidence of a 
transition at the fully convective boundary. 
On the other hand, the slope of the upper envelope of the period-mass 
relation changes sign around 0.55\,$M_\odot$, below which period rises with decreasing mass. 
\end{abstract}


\voffset=-0.6in
\hoffset=0.2in

\section{Introduction}

Of the readily observable properties of stars,
the rotation rate is one which evolves significantly on the
main sequence: intermediate- and low-mass stars are thought to spin
down throughout their lifetimes, losing angular momentum via a
magnetised wind that is linked to their outer convection zone
\citep{kaw88,bou+97}. Measuring rotation rates for large numbers of
stars over a wide range of masses and ages is a long-standing goal in
stellar astronomy, not only to understand the physical mechanisms
driving the wind and the resulting angular momentum loss, but also to
calibrate the relationship between period \prot, age $t$ and stellar
mass $M$, enabling age estimates to be made for individual stars
\citep{kaw89,bar03}. However, period measurements for
main-sequence stars with well-determined ages remain scarce,
particularly at low masses, and gyro-chronological ages are thus
restricted to a limited range of masses and ages and remain 
very uncertain \citep{bar07}.

Until the 1990s, stellar rotation rates were mainly measured from
spectroscopy, via the rotational broadening of absorption lines. These
rotational velocity measurements provided key insights, particularly
the well-known spin-down law for Sun-like stars, $\prot \propto
\sqrt{t}$ \citep{sku72}. However, these measurements yielded only
model-dependent constraints on the rotation rate, and were limited to
relatively fast rotators. Using modern wide-field detectors, it is
possible to measure \prot\ directly, by monitoring the brightness of
large numbers of stars simultaneously, and detecting quasi-periodic
brightness variations, which arise as magnetically active regions on
the star's surface rotate in and out of view. Over the past decade,
open cluster surveys have provided thousands of measurements for low
mass stars with ages up to $\sim 650$\,Myr, an overview of which can
be found in \cite{irw+09}. Notable, more recent additions to the
literature on rotation of early main-sequence low-mass stars include
studies of M37 \citep{mei+09} and Coma Berenices
\citep{col+09}. Together, these data have provided a relatively
complete, but complex picture of rotational evolution on the pre-main
sequence. In turn, this has led to renewed efforts to develop models which can
describe, or even better explain, the observations over the full
sub-solar mass range \citep{bar+10,bar10,rei+12}. 

Period measurements for older stars remain very scarce, because they
rotate more slowly and are less active than their younger
counterparts, making it is very difficult to detect their rotational
modulation from the ground. Notable exceptions include $71$ main
sequence F, G and K stars observed as part of the Mount Wilson HK
project \citep{bar03}, $1727$ mid-F to mid-K stars observed by the {\it
  CoRoT}\ satellite \citep{aff+12}, and 41 low mass
($0.1$--$0.3\,M_\odot$) stars from the \mearth\ survey
\citep{irw+11}. The \kepler\ space mission \citep{bor+10} now offers a
unique opportunity to measure rotation periods even for slowly
rotating, moderately active stars, thanks to its 
superior precision and long baseline. A previous study by \cite{har+12}
measured rotation periods for 265 stars with $T_{\rm eff} \le 5200\,$\,K and $\log g \ge  4.0$\,dex 
observed by \kepler\ for 1--2 quarters through the Cycle 1 Guest Observer program.

The few open clusters included
in \kepler's field-of-view are particularly important, since their
ages can be estimated relatively well, and a small sample of periods
has already been published for $71$ members of NGC\,6811
\citep{mei+11}. \kepler\ also observed tens of thousands of
field stars, which can yield period measurements. Although they lack
individual age estimates, they provide a global picture of stellar
spin across our Galactic neighbourhood, and can be used to constrain
the period-mass-age relation in a statistical sense. In this paper, we
focus specifically on the \kepler\ M-dwarfs with mass range
($0.3$--$0.55\,M_\odot$) where there are extremely few previous period
determinations for main-sequence objects.

Standard approaches to period detection in light curves are based on
Fourier decomposition or, for irregularly sampled data, least-squares
fitting of sinusoidal models and variants thereof \citep{sca82,zec+09}. 
However, typical stellar light curves 
are neither sinusoidal nor strictly periodic, probably because of the clumpy and
time-evolving nature of the underlying active region
distribution. Residual instrumental systematics are often present as well.

These effects can all lead to a complex periodogram structure, with
spurious peaks from jumps and long term systematics, and multiple or split 
peaks from spot evolution or differential rotation. It is therefore challenging
to determine which peak corresponds to the rotation period, 
without a priori knowledge of the range of rotation periods expected. 
Consequently, Fourier-domain methods are not always the
best suited to make the most of \kepler's many thousands of spot
modulated light curves, which display a wide range of rotation periods. 
We present an alternative approach based on the autocorrelation function (ACF) of the
light curves. To our knowledge, this is the first time that the ACF is
used as the primary tool to detect stellar rotation periods, although
\cite{aff+12} used it as a secondary verification tool.

In Section~\ref{sec:ACF_intro} we introduce the
autocorrelation function (ACF) as a robust method for period detection
in time series data. The results for the \kepler\ M-dwarf sample
are shown in Section~\ref{sec:app_kep}, and discussed in detail in Section~\ref{sec:disc}.
Section~\ref{sec:conclusions} summarises our conclusions and outlines plans for future work. 

\section{The ACF method} 
\label{sec:ACF_intro}

In signal processing, the ACF takes the standard form
\begin{equation}
  r_k = \frac{\sum_{i=1}^{N-k}  \left(x_i - \bar{x}  \right)
    \left(x_{i+k} - \bar{x}  \right) } 
{\sum_{i=1}^{N}  \left(x_i - \bar{x}  \right)^2}\, ,
\label{eq: acf_eq}
\end{equation}
{\citep[see e.g.][]{shst10} where $r_k$ is the autocorrelation coefficient at lag $k$, for time
series $x_i$ ($i = 1, \dots, N$). Each lag $k$
corresponds to $\tau_k = k \Delta t$, where $\Delta t$ is
the cadence. In our implementation, the 
light curves are median normalised before the ACF is computed, 
and we only search for periods less than half the length of the dataset,
i.e. $k$ \textless\ $N / 2$.

We compare the ACF method to the method most commonly
used to search for rotation periods in stellar light curves, namely
least squares fitting of sinusoids over a grid of trial periods
\citep{irw+06,zec+09}. The amplitude, phase and zero-point of the
sinusoid are free to vary. The sine-fitting periodogram is expressed in
terms of the statistic 
\begin{equation}
  S = \left(\chi_{0}^{2} - \chi^{2}\right) / \chi_{0}^{2}\, ,
\end{equation}
where $\chi^{2}_{0}$ is the reduced chi-squared of the light curve
with respect to a constant value, and $\chi^{2}$ is the reduced
chi-squared with respect to the best-fit sinusoid. This method is
described in more detail in \cite{mcq+11}. 

Fig.~\ref{fig:acf_examples1} shows two synthetic time-series
curves, together with their ACFs and least-squares sine 
curve fitting periodograms \citep{zec+09}. The left column shows a 
strictly periodic signal, for which the periodogram displays a clear 
pronounced peak. The ACF displays an oscillatory behaviour, with regularly
spaced peaks located at multiples of the period. The
amplitude of these peaks decays gradually because of the
definite duration of the time-series. The right column
shows the effect of introducing correlated noise to the signal
in the left hand column. 

\begin{figure}
  \centering
  \includegraphics[width=\linewidth]{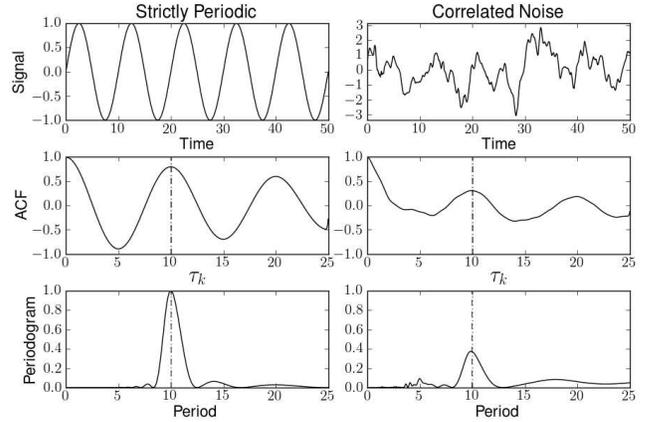}
  \caption{Simple synthetic signal of amplitude 1 (top row), and corresponding 
  ACF (centre row) and periodogram (bottom row). The right hand column
  shows the effect of introducing noise of amplitude 0.9, correlated on a 
  2 time unit timescale. On the ACF and the periodogram panels, the 
  input period used to generate the signal is shown as a 
  vertical dotted line, detected period from each corresponding
  method is marked as the over-plotted dashed line.}
  \label{fig:acf_examples1}
\end{figure}

\begin{figure*}
  \centering
  \includegraphics[width=\linewidth]{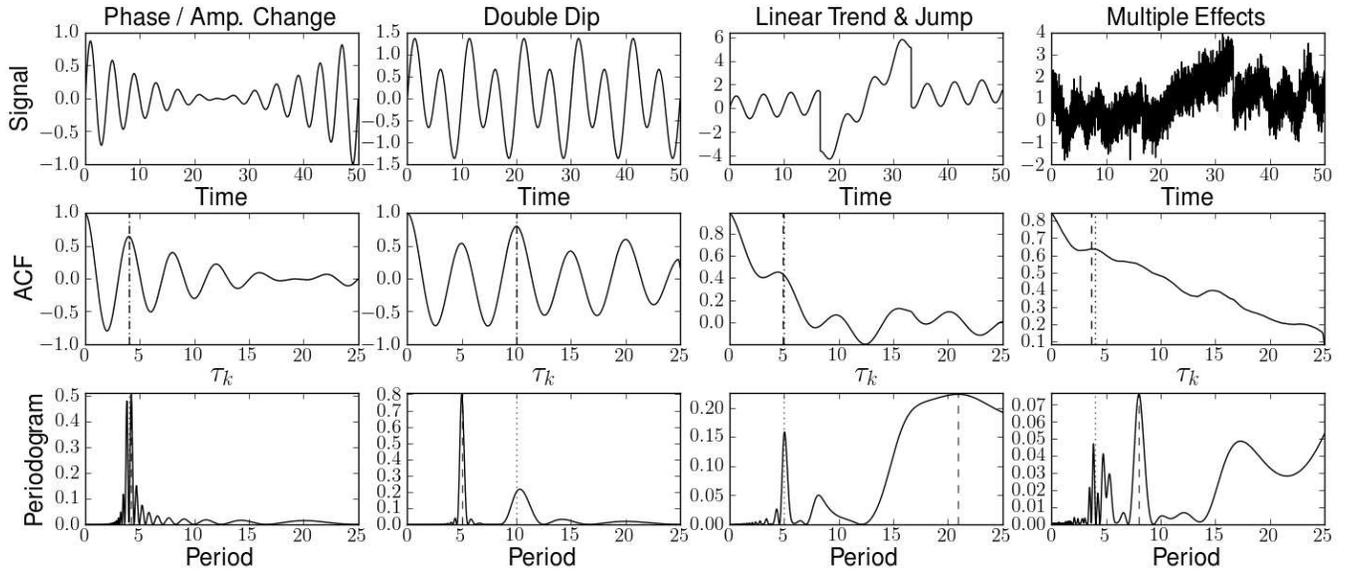}
  \caption{Synthetic examples showing the effects of varying signal 
  phase and amplitude and injecting noise and systematics into the 
  sine curves (top row), on the ACFs (middle row) and periodograms 
  (bottom row). The `Multiple Effects' signal comprises of the `Phase/Amp. Change' 
  signal from the first column, with injected white noise, correlated noise, 
  a linear trend and a jump. On the ACF and the periodogram panels, the 
  input period used to generate the signal is shown as a vertical 
  dotted line, detected period from each corresponding method 
  is marked as the dashed line, overlaying the dotted line in most cases. }
  \label{fig:acf_examples2}
\end{figure*}

In Fig.~\ref{fig:acf_examples2} we demonstrate how the ACF
and the periodogram are affected by varying signal phase and amplitude, 
systematics effects and noise. 

In cases where the phase and amplitude of the signal vary with time, the correct
period is detected but the ACF peak amplitude varies within an envelope,
corresponding to the amplitude variation of the signal. The peak width 
and level of symmetry can also vary. In this case, the periodogram produces 
two peaks on both sides of the correct period. 

If the signal contains multiple minima and maxima per period,
as can occur for spotted stars with more than one dominant active
region, the ACF often shows alternating low and high ACF peaks, due to
a partial correlation between the sets of maxima or minima.
Our algorithm is built to identify these cases 
(see Section~\ref{sec:meas}), and in this case selected the right period. 
On the other hand, the periodogram picked half
the right period.

A jump or long term trend in the signal 
introduces a long term trend in the ACF, 
and since this can take many shapes, it is important to look at the local
variations in peak height when performing diagnostics on the ACF. These
long term trends introduce a long-period peak in the periodogram, 
which can lead to a wrong identification of long periodicity. 
We therefore expect the ACF method to be more reliable in these
cases.

Noise and systematics are present in many
stellar light curves displaying rotational modulation, 
and must be accounted for when
attempting to determine rotation periods using the ACF. The effect of
combining these factors is shown in the right
column of Fig.~\ref{fig:acf_examples2}, which has phase and amplitude modulation,
white and correlated noise, a linear trend and a jump.

In summary, because the ACF measures only the degree of
self-similarity of the light curve at a given time lag, the period
remains detectable even when the amplitude and phase of the
photometric modulation evolve significantly during the time-span of
the observations. The ACF method is also capable of
producing robust results in cases with residual instrumental 
systematics, because correlated noise, long-term trends and
discontinuities give rise to monotonic trends in the ACF, on top
of which we are able to identify the local maxima. Therefore, the
ACF method is expected to be more robust to active region 
evolution than the periodogram, which implicitly assumes a stable, 
sinusoidal signal. The analysis of the real \kepler\ data supports 
this conclusion, as detailed in Section~\ref{sec:comp_pgram}.
 
\subsection{Measuring periods from the ACF}
\label{sec:meas}
 
The period measurement involves three steps: identifying peaks in the
ACF, selecting the peak associated with the mean rotation period, if any, and
evaluating the uncertainty on the period.

The presence of high-frequency noise in the light curves leads to
numerous local extrema in the ACF. Therefore, we first smoothed the
ACF by convolving it with a Gaussian kernel, of window size of 56
lags and a full width at half-maximum ({\sc FWHM}) of 18 lags. These values
were tuned to provide the best compromise between reducing noise and
maintaining ACF signal, without prior knowledge of the period. 
See Fig.~\ref{fig:smoothing} for an example of the effects of the smoothing treatment. 
We then identify local extrema in the smoothed ACF, defined as locations where the
gradient changes sign.

\begin{figure}
  \centering
  \includegraphics[width=\linewidth]{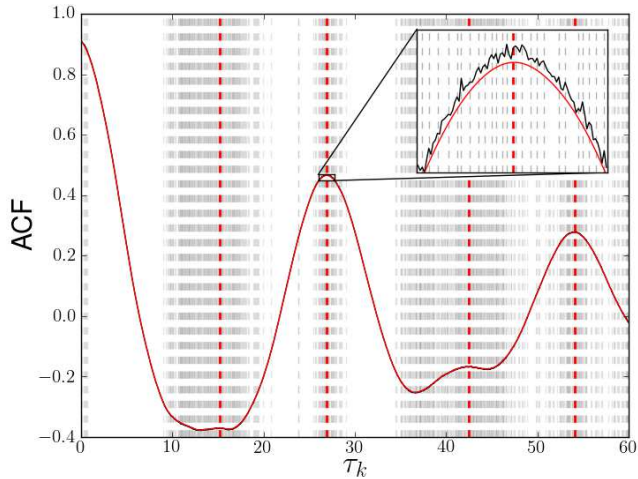}
  \caption{Example of an unsmoothed ACF (black curve) and 
  corresponding smoothed curve (red). The many small peaks detected in the 
  unsmoothed curve are shown as grey vertical dashed lines and the 
  peaks detected from the smoothed version are shown as red vertical dashed lines.
  The inset plot shows a section of the ACF in detail.}
  \label{fig:smoothing}
\end{figure}

If the light curve contains a clear rotational modulation signal, 
this process yields a series of clear, regularly spaced peaks of gradually decreasing height,
as seen in Fig.~\ref{fig:acf_examples1}. 
The first peak corresponds to the interval between patterns in the light curve, 
which evolve gradually, but are clearly repeated, and is thus identified 
as the rotation period. Some of the light curves contain long term 
trends and discontinuities, as a result of imperfections in the systematics 
correction. These introduce power at low frequencies, and thus affect the 
behaviour of the ACF for large lags, but the first ACF peak still 
corresponds to the correct period (as identified by visual examination 
of the light curve).

In the remaining steps, we make use of the height of the ACF peaks. 
However, correlated noise and residual systematics can introduce 
underlying long term trends, which mean the absolute peak height is no longer a good 
diagnostic. To mitigate this effect, we measure the height
of each peak relative to the two adjacent minima, and adopt the
mean of the two measurements as the `local height' of the peak, denoted by $h_{\rm P}$.
Since the ACF values range from -1 to 1, $h_{\rm P}$ has a positive value, of maximum 2. 

If a star has two dominant active regions located on opposite
hemispheres, each causes a series of dips in the light
curve, approximately in anti-phase with each other. This gives rise to
a partial correlation at half the mean rotation period, leading to a peak in
the ACF at $\tau_k = P/2$. However, the peak at $P/2$ is typically
smaller than that at $P$, because the two active regions do not generally give 
rise to identical dips. We therefore adopt the following condition: if $h_{\rm P}$ of the 
second peak is greater than that of the first, the second peak is selected instead. 
The right panel of Fig.~\ref{fig:synth_ex} and the top left panel of Fig.~\ref{fig:kep_ex} show
a synthetic and real example of this effect, where this method has detected 
the correct peak.

In some cases, correlated noise and residual systematics 
produce an underlying slope at small $\tau_k $, causing a shift the 
position of the first peak associated with the rotation period. This occurs 
because a peak on a slope will have its maxima shifted in the direction 
of increasing gradient, as seen in the left panel of Fig.~\ref{fig:synth_ex}. 
To avoid this bias, a more robust period measurement 
is obtained using the median of the intervals $\Delta \tau_k$ between 
consecutive ACF peaks associated with the rotation period. To avoid 
selection of erroneous peaks, only those located at or close to 
(within 20\%) of integer multiples of $\tau_k$ of the selected 
peak are used.  If there are several locations selected 
around each peak (i.e. the smoothing has failed to remove all erroneous 
peaks), the peak selection only occurs after a gap of $\Delta \tau_k > 0.3\ \tau_k$.
This ensures only 1 data point per peak is used in the period measurement.
Since the accuracy of the peak positions can decrease for very large $\tau_k$, a maximum 
of 10 peaks was selected for measurement of the median period and uncertainty.

We define the period uncertainty as the scatter of the $\Delta \tau_k$. Specifically, we used
\begin{equation}
\sigma_P = \frac{1.483 \times \rm{MAD}}{\sqrt{N-1}}
\label{eq: mad}
\end{equation}
where $N$ is the number of peaks, and {\sc MAD} is the median of the
absolute deviations from the median $\Delta \tau_k$. This
`{\sc MAD}-estimated scatter' is equivalent to the standard deviation for a
Gaussian distribution, but is more robust to outliers.    
In cases where we identified only one peak matching
the selection criteria, we adopt the peak position as the period
and the half width at half local peak height as the period uncertainty.

\subsection{Tests with Realistic Light Curves with Injected Simulated Stellar Modulation}

To demonstrate the robust nature of the ACF, 
we present a selection of synthetic examples with known input parameters.
A more detailed evaluation of the method over the full range of parameter
space will be described in future work.

\begin{figure*}
  \centering
  \includegraphics[width=0.8\linewidth]{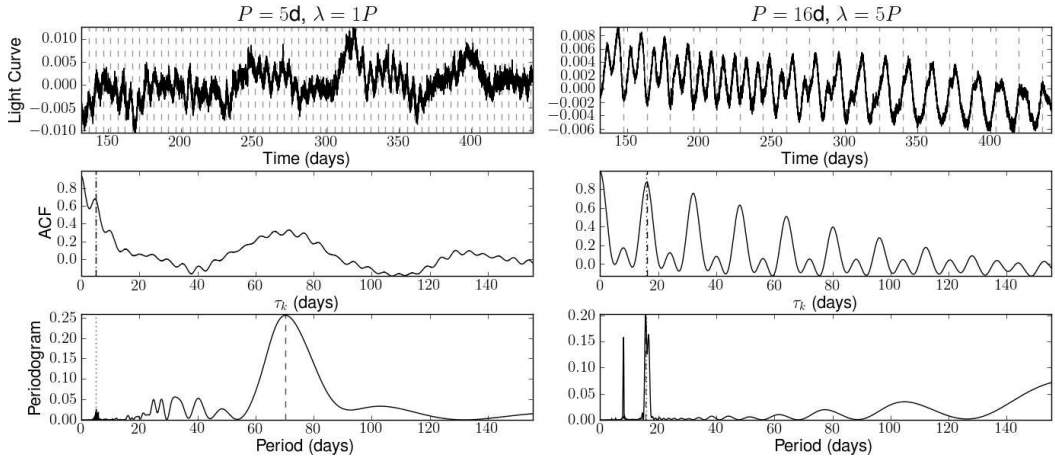}
  \caption{Examples of synthetic light curves and corresponding ACFs 
  and periodograms for different input parameters, indicated in the plot titles. 
  In the left panel the spot distribution was selected to give a double-dip light curve.
  The dashed lines on the light curve plots indicate intervals at the period detected by the ACF. 
  On the ACF and periodogram plots, the input period is marked by a dotted line and the 
  detected period is marked by a dashed line.}
  \label{fig:synth_ex}
\end{figure*}

To generate synthetic light curves we used a simple 
spot model code, described in \cite{aig+12}, which takes the following
input parameters: number of spots, light curve amplitude, characteristic spot 
half-life ($\lambda$), light curve duration, fractional differential rotation, inclination and time sampling.

The observations of \cite{jac+12} show that activity is better explained by
a large number of small spots, than a small number of large spots. They suggest
a spot number of 2500-5000, however, since they observe younger stars
we opt for a smaller spot number of 200. This also increases the speed of
synthetic light curve generation. We selected an amplitude of 1\% and an inclination
angle of $75\ensuremath{^\circ}$. The time sampling and duration match
that of the \kepler\ Q1-4 data.

Since the effects of the noise and residual systematics of the \kepler\ data 
are an important factor when analysing the ACF, we introduce these into the synthetic light
curves. A random selection of the \kepler\ M-dwarf light curves were visually examined
and 10 were selected which appeared to contain only noise and systematic effects, 
with negligible intrinsic stellar signal. For each synthetic output, we created 
10 light curves with realistic noise and systematics by adding each of the
selected quiet \kepler\ light curves.

The two synthetic examples in Fig.~\ref{fig:synth_ex} demonstrate
the ability of the ACF to recover the input period from quite difficult cases, 
where the periodogram failed to identify the right period.

The many synthetic cases we tried led us to believe that the 
ACF method and the periodogram are less reliable for light curves with very
fast spot evolution ($\lambda\,\textless\,1$\,period), short 
periods ($\textless\,7$\,days) and strong systematics. This
is caused by the steep initial slope in the ACF, combined with the smoothing
algorithm, which masks the peaks. Similarly for fast evolution, strong systematics 
and long periods ($\textgreater\,35$\,days), the ACF peaks are too strongly affected
by underlying trends. The periodogram suffers mainly from spurious peaks at 
long periods.

As the spot half-life is increased to 1 period, the periodic signal
strength increases, and by $\lambda$ = 5 periods, the majority of periods are correctly 
detected, across the range of input values. It should be noted that these synthetic 
examples were designed to test the limits of the ACF and therefore
all have a low signal to noise. A recent study of a sample of Sun-like stars 
observed by CoRoT \citep{mos+09} found that spot lifetime of the order 
of the rotation period are not atypical for Sun-like stars. On the other hand, 
the spot lifetime is also known to increase with increased activity level \citep{hahe94}.

\subsection{Amplitude of variability}
 
To estimate the average amplitude of the periodic signal, we use a
modified version of the range statistic $R_{\rm var}$\ defined by
\cite{bas+10,bas+11} and \cite{mcq+11}, which is the interval between the
$5^{\rm{th}}$ and $95^{\rm{th}}$ percentiles of the light curve. This
rank-order approach enables us to quantify the amplitude of the
variability without specifying a particular model for it, and
excluding the extreme values minimises the sensitivity to outliers and
high-frequency noise. We divide the light curve into segments, whose
duration is equal to the detected period, measure $R_{\rm var}$\
separately in each of them, and define $R_{\rm per}$\ as the average of
the $R_{\rm var}$\ measurements obtained in this way. In most cases,
$R_{\rm per}$\  is very similar to the $R_{\rm{var}}$\ value
for the whole light curve.

\section{Application to Kepler Data M-Dwarfs}
\label{sec:app_kep}

\subsection{Target Selection}
\label{sec:tar_sel}

Following \cite{cia+11}, we selected M-dwarfs
among the \kepler\ targets based on the effective temperature $T_{\rm
  eff}$ and surface gravity $\log g$ reported in the \kepler\ Input
Catalog, or KIC \citep{bro+11}.  These parameters were estimated using
Bayesian posterior probability maximisation \citep{bro+11}, matching
observed colours, estimated from Sloan $g$, $r$, $i$, $z$ filters,
2MASS $JHK$, and D51 (510 nm), to the stellar atmosphere models of
\cite{ck04}, resulting in typical uncertainties of $200\,$K for
$T_{\rm eff}$ and $0.5$\,dex for $ \log g$, respectively. These are
the typical errors for each parameter, but the actual errors may
vary by a small amount between stars of different magnitude and
spectral type.  For a more detailed discussion of the KIC parameters,
see \cite{bro+11}, \cite{bat+10} and \cite{ver+11}. There is increasing 
evidence that the KIC values for the radii and $T_{\rm eff}$ are overestimated 
\citep[e.g.][]{man+12, ver+11, mui+12}, which means the values of 
$T_{\rm eff}$ should be interpreted with care.

This study focusses on M-dwarfs stars, which were selected following
\cite{cia+11} as having $T_{\rm eff} \le 4000$ K and $ \log g \ge
4.0$ dex, manually adding 25 known M-dwarfs \citep{cia+11} with missing KIC
parameters. Table~\ref{tab:numbers} summarises the number of stars
considered at each stage of the study.

\begin{table}
  \caption{Number of objects included at each stage of the study. The
    column labelled `All dM' includes both M-dwarfs selected from the KIC parameters and
    the previously known M-dwarfs without KIC
    parameters. The totals in the second column include the known
    EBs and planet candidate host stars, which are also listed
    separately in the last two columns. The possible giants and double-period
    stars were removed from the final results reported in
    Tables~\protect\ref{tab:prot} \& \protect\ref{tab:no_prot}.
    Note: removed 4 contact EBs from Q1-4 EB selection.}
  \label{tab:numbers}
  \centering
  \begin{tabular}{lcccc} 
    \hline
    Stage   &    All dM  &   EBs & Pl.\ host \\
    \hline\hline
    KIC selection                    &    2937     &          15    &      57   \\ 
    Q1-4 light curves            &    2483     &        9    &      51   \\ 
    Period detected               &    1730      &        7     &      42   \\
    Possible giants               &     121       &            0     &      0     \\
    Double-period                &     39         &            0     &      0      \\
    Binaries or pulsators      &     112       &             -      &      0       \\
    Final periodic total         &     1570     &           7      &      42    \\
   \hline
 \end{tabular}
\end{table}

\subsection{Obtaining and pre-processing the light curves} 

This study is based on public release 14, which was available 
when the present analysis was performed. The release 
included data from Quarter 1 (Q1) to Quarter 4 (Q4) of \kepler\
observations, which took place over $\sim 310$ days between May
$13^{\rm th}$ 2009 and March $19^{\rm th}$ 2010. Approximately
$156\,000$ targets were observed during this time, most with a cadence
of $29.42$\,minutes. These data are publicly available and were downloaded from
the \kepler\ mission archive ({\tt http://archive.stsci.edu/kepler}) at
the Space Telescope Science Institute (STScI). 

Public release 14 included a re-processing of Q1--\,Q4 with a new 
version of the presearch data conditioning (PDC) 
pipeline known as PDCMAP. The purpose of the PDC is to remove the majority
of instrumental glitches and systematic trends. In contrast with earlier
versions, the new PDC-MAP uses a Bayesian approach to do so while
retaining most real (astrophysical) variability \citep{smi+12,stu+12}.
It works very well for Q3 and Q4, but some artefacts remain in some
light curves in Q1 and Q2. Improving the correction of instrumental
effects further remains an important goal to enable the full
exploitation of \kepler's potential in terms of stellar
astrophysics. However, in the vast majority of the cases we have
examined, the residual artefacts in the PDC-MAP data do not prevent
the detection of rotational modulation, so we consider it suitable for
the present work.

An additional 5 quarters of PDC-MAP data were made publicly available
a few weeks before the submission of this manuscript. We are planning
to analyse these in the near future, in the hope that it will improve
our sensitivity to long periods, but we do not expect it to affect the
main conclusions of this paper. 

The ACF calculation requires the light curves to be regularly sampled
and normalised to zero. We divided the flux in each quarter by its
median and subtracted unity. Gaps in the light curve longer than the
\kepler\ long cadence were filled using linear interpolation with added
white Gaussian noise. This noise level was estimated using the
variance of the residuals following subtraction of a smoothed version of the flux. 
To smooth the flux we applied an iterative nonlinear filter which consists 
of a median filter followed by a boxcar filter, both with 11-point windows, 
with iterative 3-sigma clipping of outliers.

This method of gap filling and quarter stitching was only
found to introduce spurious jumps in light curves for which the systematics 
correction had failed, and were therefore already problematic. In the vast
majority of cases, the amplitude of M-dwarf activity mean that any quarter stitching 
effects have a negligible impact on the period detection.

Our sample also includes a number of previously
identified eclipsing binaries and planetary transit candidates
\citep{prs+11,bat+12}. In those cases we cut out and interpolated over
the eclipses/transits before performing the rotation period search.

\subsection{Establishing confidence in the period detection}

Given the scientific importance and manageable
size of the sample under study here, we opted to perform a visual
examination of all the light curves in order to verify the period
detected by the ACF. The light curves were compared to 
dashed vertical lines at intervals of the detected period (see e.g. Fig.~\ref{fig:kep_ex}).
In order to classify the detection as valid, features in the light curve must be present at
intervals matching the dashed lines, across several periods, preferably in more than 
1 quarter.

\begin{figure*}
  \centering
  \raisebox{10pt}{\includegraphics[width=0.8\linewidth]{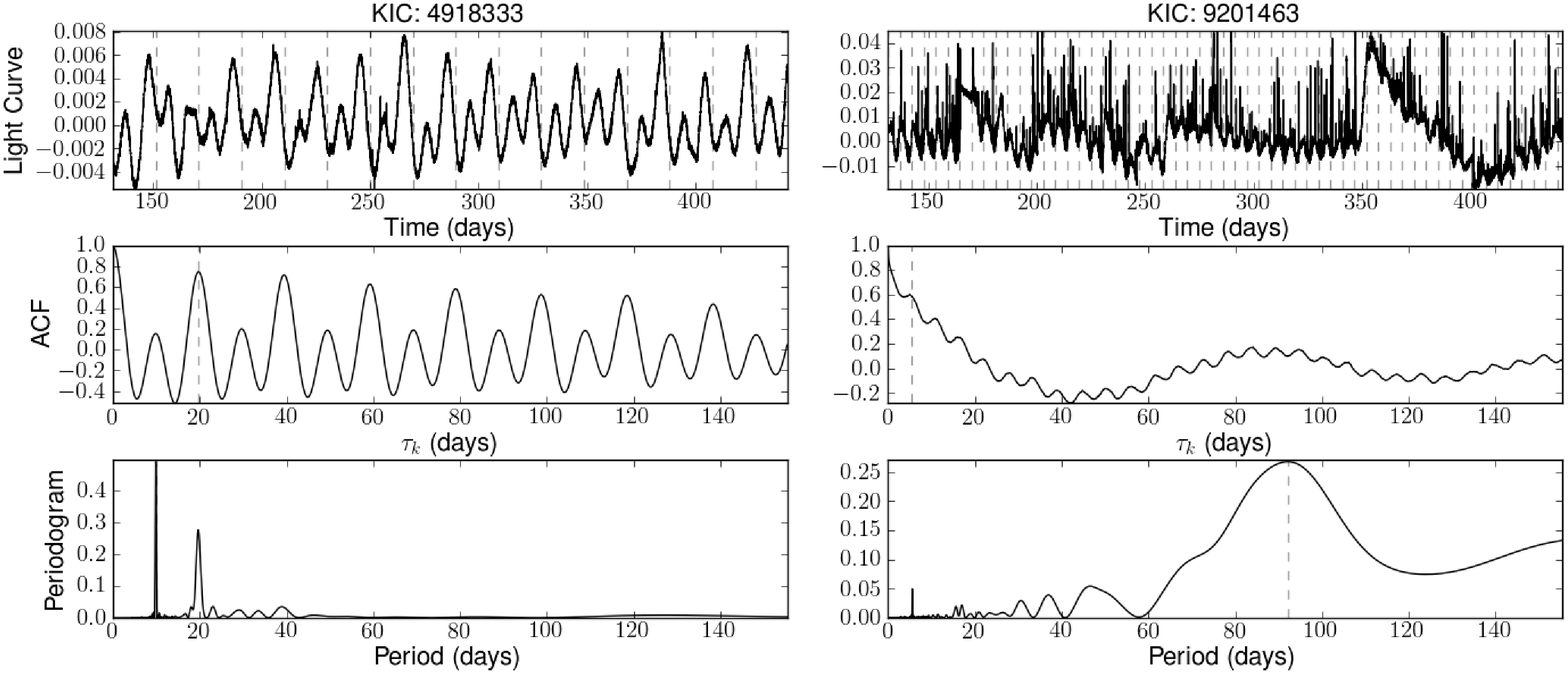}}
  \includegraphics[width=0.8\linewidth]{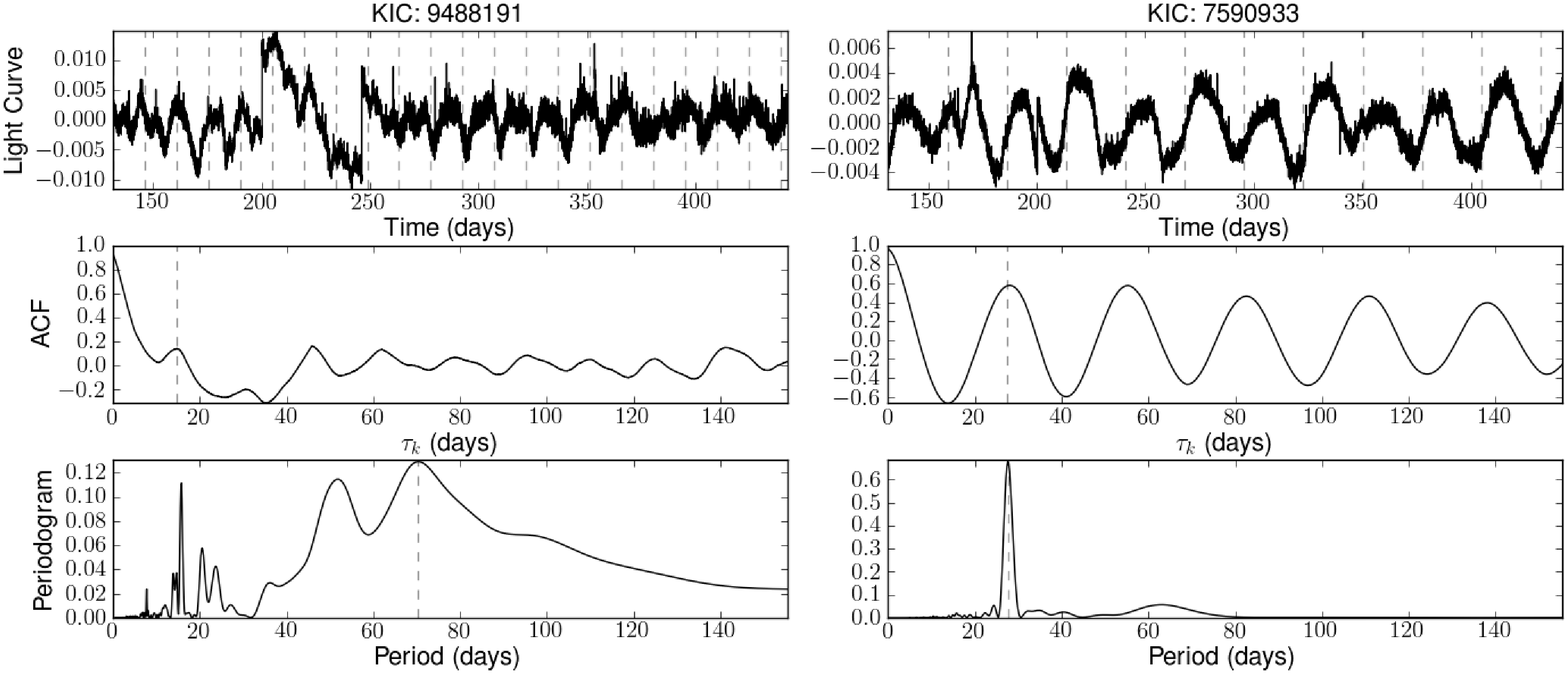}
  \caption{Examples of \kepler\ light curves and corresponding ACFs. The dashed lines 
  on the light curve plots indicate intervals at the period detected by the ACF 
  (dashed line on ACF plot). KIC\,4918333 shows two active regions, creating
  a double dip effect. The correct period was automatically detected by selecting the second
  (higher) ACF peak. KIC\,9201463 shows that the ACF is robust against flares and 
  significant systematics. KIC\,9488191 contains significant residual systematics and the effect
  on the ACF is apparent, however, the correct period is still detected. KIC\,7590933 shows an
  example which has been marked as 'possible harmonic' or `HM' in Table~\ref{tab:prot}, 
  since it is not clear whether the rotation period corresponds to the first or second ACF peak, 
  even though the ACF peak height indicates it should be the first (see text for further discussion).}
  \label{fig:kep_ex}
\end{figure*}

We paid particular attention to the question of whether the
detected period was clearly the rotation period, or could be $P/2$, 
as described in Section~\ref{sec:meas}. In 73\% of 
light curves we identified features which could be `tracked' 
visually, i.e. a particular shape of spot crossing
that repeats throughout the light curve. When there are
more than one such sets of features, and they evolve gradually
relative to each other, as shown for example in the top left panel of Fig.~\ref{fig:kep_ex}, 
the rotation period is very clear. When only one such set is visible and the first ACF peak
is not higher than the second (e.g. the bottom right panel of Fig.~\ref{fig:kep_ex}), we
cannot be so certain, but the simplest explanation is nonetheless that
the detected period is correct. In around 1\% of the cases where we
report a period, there is very tentative evidence in the light curve
that the detected period is a harmonic (i.e. $P/2$), but we could not be certain,
so we simply flagged the corresponding objects as `possible harmonics'. 
Such cases may be resolved once additional quarters of \kepler\ data are incorporated
into the analysis. We note that these cases are not numerous enough to
explain the two sequences seen in the period-temperature diagram (Fig.~\ref{fig:per_teff}), and are
spread across the entire period range. 

In 3.6\% (57) of the periodic cases an incorrect peak was identified.
This was most frequently a result of noise introducing extra peaks, or very 
large residual systematics changing the relative peak heights. These cases were
manually corrected to select peaks corresponding to the period identified by eye.

In the future, we plan to apply the same analysis
to a much larger sample of \kepler\ targets (including F, G and K
dwarfs), for which an automated detection method must be developed.
The present, visually inspected subset will then prove
valuable, to validate any threshold to be applied automatically to a
larger sample. Here we merely note that, in the present study 
using 10 months of data, it is possible to recover 91\% of our 
detections, at the cost of a false alarm rate of 10\%, by selecting objects with 
\begin{equation}
h_{\rm P} > {\rm MAX}\left(0.15, \frac{\sigma_{\rm P}}{51\,{\rm days}}\right).
\end{equation}

\subsection{Results}
\label{sec:acf_imp}
 
\begin{table*}
  \caption{M-dwarfs with detected rotation periods.
  This table is available in its entirety, in a machine-readable form
   in the online supplementary material, or at {\tt http://www.physics.ox.ac.uk/StellarRotation}. A portion is shown here for guidance 
  regarding its form and content. $T_{\rm eff}$ and $\log g$ are from the KIC and $M$ was
    derived from $T_{\rm eff}$ using the 600 Myr isochrone of
    \protect\citet{bar+98}. The average amplitude of variability per period bin of the light curve, $R_{\rm per}$,
    is included. The meaning of the flags are: `EB': known
    eclipsing binary \protect\citep{prs+11}; `PL': planet host
    candidate \protect\citep{bat+12}; `PB': ultra-stable periodic
    behaviour, indicating possible binary, pulsator or young object; `HM': the
    reported period may be a harmonic of the true period; `NF': no flag. }
  \label{tab:prot}
  \centering
  \begin{tabular}{ccccccccc}
\hline
    KIC & $T_{\rm eff}$ & $\log g$ & $M$ & $\prot$ & $\sigma_{\rm  P}$ & $R_{\rm per}$ & Flag \\
         & (K) & (g/cm$^3$) & ($M_{\odot}$) & (days) & (days) & (mmag)
         & \\ \hline\hline
1162635&3899&4.62&0.5037&15.509&0.064&10.7&NF\\
1430893&3956&4.41&0.5260&17.144&0.046&10.4&NF\\
1572802&3990&4.48&0.5394&0.368&0.000&74.8&PB\\
1721911&3833&4.58&0.4781&28.403&0.394&3.9&NF\\
1866535&3878&4.50&0.4955&25.052&0.136&4.0&NF\\ \hline
  \end{tabular}
\end{table*}

We applied the ACF method and visual inspection steps described in the
previous section to the \kepler\ light curves of the objects selected
as likely M-dwarfs. The results are reported in Tables~\ref{tab:prot}
to \ref{tab:no_prot}. The full machine-readable tables are available
in the online supplementary material, or with plots of every
light curve, its ACF, and sine-fitting periodogram at the URL
{\tt http://www.physics.ox.ac.uk/StellarRotation}. 
Table~\ref{tab:prot} reports all our period measurements,
except for two groups of objects, listed separately in
Tables~\ref{tab:giants} and \ref{tab:double}, which were excluded from
the sample for reasons detailed below. Table~\ref{tab:no_prot} lists
all the objects which passed our target selection criteria, but for
which no period was detected. The final number of 
likely M-dwarfs with detected rotation periods is 1570 
(out of the 2483 light curves we analysed, or 2362 if we 
discount objects later found to be giants).

\subsection{Excluding non-rotators}

Rotational modulation of star spots is not the only cause of periodic,
or quasi-periodic, variability in stellar light curves. The visual
examination stage was therefore important in identifying groups of
stars whose variability did not appear to be caused by rotation, or by
rotation alone.

The \kepler\ light curves have already been searched extensively for
planetary transits \citep{bor+11,bat+12} and stellar eclipses
\citep{prs+11}. We included these objects in our sample, after removing
the transits and eclipses, as described earlier. Measuring the photometric 
rotation periods of  known binaries is useful to check for differences 
between the rotational properties of close binaries and of apparently single stars. The
rotation of planet-host stars is an important topic in itself, which
is beyond the scope of the present paper and will be investigated in
detail in a forthcoming publication.

\begin{table}
  \caption{Objects identified as likely giants from their light curves.
  This table is available in its entirety, in a machine-readable form
   in the online supplementary material, or at {\tt http://www.physics.ox.ac.uk/StellarRotation}. 
   A portion is shown here for guidance 
  regarding its form and content.}
  \label{tab:giants}
  \centering
  \begin{tabular}{ccccc}
\hline
    KIC & $M$ &  $\log g$ & $T_{\rm eff}$ & $R_{\rm var}$ \\
         & ($M_{\odot}$) & (g/cm$^3$)  & (K) & (mmag) \\ \hline\hline
1026895&0.5343&4.53&3977&8.99\\
1160867&0.4474&4.57&3753&8.29\\
1431599&0.5068&4.5&3907&12.68\\
1576043&0.4889&4.52&3861&7.76\\
2010137&0.4881&4.48&3859&7.01\\
\hline
  \end{tabular}
\end{table}

\begin{figure}
  \centering
  \includegraphics[width=\linewidth]{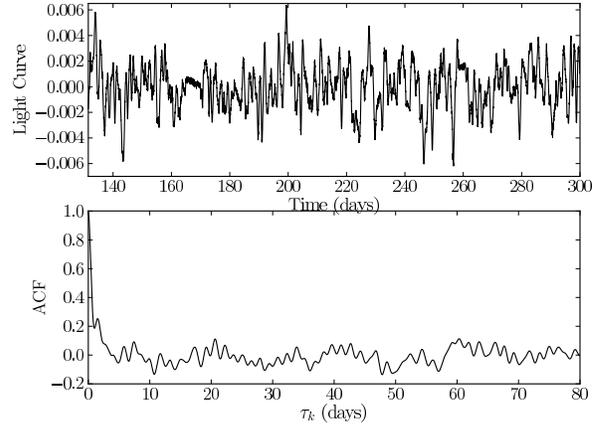}
  \caption{KIC\,2010137 is an example of a star which we identified as
    a likely giant based on its light curve, which shows stochastic
    variability with a clear, dominant time-scale, but no
    repeatability, and does not resemble the signal expected from
    rotation.}
  \label{fig:odd_example}
\end{figure}

One distinct group of 121 stars show a clear ACF peak at relatively
short $\tau_k$ ($\sim 1$--$12$ days), but there are few additional
peaks at integer multiples of the first, and the light curve appears
stochastic rather than truly periodic. Such an example is shown in
Fig.\,\ref{fig:odd_example}. We initially thought that these objects
may be rapid rotators with very rapidly evolving active regions, or a
hitherto unidentified type of pulsating star. We examined their KIC
parameters, and noted that they appear redder, and have lower proper
motions, than the rest of the sample. We therefore concluded that they
are likely to be giants contaminating our sample. Indeed, if we apply
the $J-H < 0.75$ cut advocated by \cite{cia+11},
it removes 117 stars from the M-dwarf sample, of which 103 belong to this group of
objects displaying stochastic behaviour. We therefore concluded that
these objects are likely giants, removed them from the M-dwarf sample
and list them separately in Table~\ref{tab:giants}.

A further 39 objects display evidence for two, very distinct periods
in their light curves, with one period several times longer than the
other. This could potentially result from the light of
two different variable stars being included in a single photometric aperture,
which may be revealed by examination of the pixel level data.
Since the nature of the periodicity is unknown, and the determination
of the KIC parameters used to select these objects as M-dwarfs 
could be affected by the presence of a close companion, we exclude these
objects from our sample, and list them separately in
Table~\ref{tab:double}. Note we do not report periods for them,
because the presence of two distinct signals in their light curves
makes the identification of either period more challenging.

\begin{figure}
  \centering
  \includegraphics[width=\linewidth]{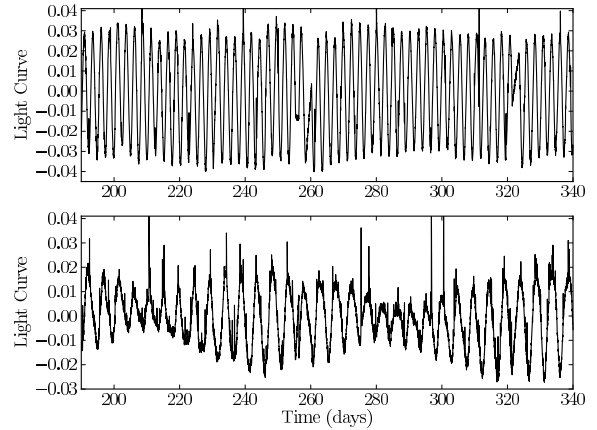}
  \caption{Examples of the objects classified as `PB'. KIC\,5516671 (top) shows
  a very stable light curve and KIC\,3103752 (bottom) shows the `beat pattern' discussed 
  in the text.}
  \label{fig:pb_plot}
\end{figure}

\begin{table}
  \caption{Objects with two distinct periods in their light curves. 
  This table is available in its entirety, in a machine-readable form
   in the online supplementary material, or at {\tt http://www.physics.ox.ac.uk/StellarRotation}. 
   A portion is shown here for guidance 
  regarding its form and content.}
  \label{tab:double}
  \centering
  \begin{tabular}{ccccc}
\hline
    KIC & $M$ &  $\log g$ & $T_{\rm eff}$ & $R_{\rm var}$ \\
         & ($M_{\odot}$) & (g/cm$^3$)  & (K) & (mmag) \\ \hline\hline
892376&0.4704&4.47&3813&13.84\\
1569863&0.4688&4.45&3809&33.94\\
2557669&0.4585&4.36&3782&37.7\\
3646734&0.4372&4.47&3726&25.89\\
3735772&0.4414&4.48&3737&85.71\\
\hline
  \end{tabular}
\end{table}

Finally, 109 stars show unusually stable periodic behaviour, with
periods typically $< 7$\,days, and very little or no evidence of any
evolution over the full Q1--Q4 duration. Some of these also show a
beat pattern characteristic of the light curves containing two, mutually
similar periodicities. See Fig.~\ref{fig:pb_plot} for examples. 
These objects could be members of close binary
systems, where the active region pattern is stabilised over long
timescales because of the presence of a companion. Their rotation
periods and amplitudes are certainly similar to those of the known
eclipsing binaries in the sample. If they are binaries, the beat
patterns may indicate differential rotation on one of the stars, or
two slightly different rotation periods for the two components of the
binary. We are not aware of any type of main-sequence
M-dwarf that would be expected to pulsate in this period range, but we cannot 
rule out binarity or pulsation without spectroscopy. We
therefore kept these objects in Table~\ref{tab:prot}, but flagged them as 
`binaries or pulsators', indicating their period may result from a
phenomenon other than rotation.

\begin{table}
  \caption{Objects with no rotation period detection. 
  This table is available in its entirety, in a machine-readable form
   in the online supplementary material, or at {\tt http://www.physics.ox.ac.uk/StellarRotation}. 
   A portion is shown here for guidance 
  regarding its form and content.}
  \label{tab:no_prot}
  \centering
  \begin{tabular}{ccccc}
\hline
    KIC & $M$ &  $\log g$ & $T_{\rm eff}$ & $R_{\rm var}$ \\
         & ($M_{\odot}$) & (g/cm$^3$)  & (K) & (mmag) \\ \hline\hline
1160684&0.5244&4.48&3952&3.35\\
1292688&0.5146&4.86&3927&7.48\\
1569682&0.5154&4.69&3929&4.99\\
1718059&0.526&4.49&3956&3.13\\
1718071&0.5225&4.54&3947&5.77\\
\hline
  \end{tabular}
\end{table}

\subsection{Comparison of the AFC and sine-fitting periodogram}
\label{sec:comp_pgram}

We computed sine-fitting periodograms for the sample of \kepler\ M-dwarfs
using 1000 logarithmically-spaced periods between 0.1 and 155 days.

In general, similar results are obtained from
the ACF or periodogram methods, however, we consider the ACF
clearer and more reliable. A useful feature of the ACF method
is that it enables automatic identification of
cases where the detected period is a half the mean rotation period. 
An example of this is shown for
KIC\,4918333 in Fig.~\ref{fig:kep_ex}, which clearly shows that the 
mean rotation period corresponds to the second ACF peak, 
whereas the first periodogram peak is the highest.

Similarly, for KIC\,9488191 in Fig.~\ref{fig:kep_ex},
one can determine easily from the ACF that the 
rotation period corresponds to the first peak at $\sim 15$ days and not 
the highest peak in the periodogram at $\sim 70$ days. Although 
long period peaks could be excluded from the periodogram to allow
the correct peak to be selected, this would require 
prior knowledge of the possible range of rotation periods to avoid removal of genuine long period
signatures.

A more systematic comparison between the 
ACF and periodogram results is shown in Fig.~\ref{fig:acf_sine}.
Here we make a one-to-one comparison of the periods detected 
using the ACF and the sine-fitting periodogram, for all the cases 
where the light curves were determined to show periodic variations
from visual examination.

\begin{figure}
  \centering
  \includegraphics[width=\linewidth]{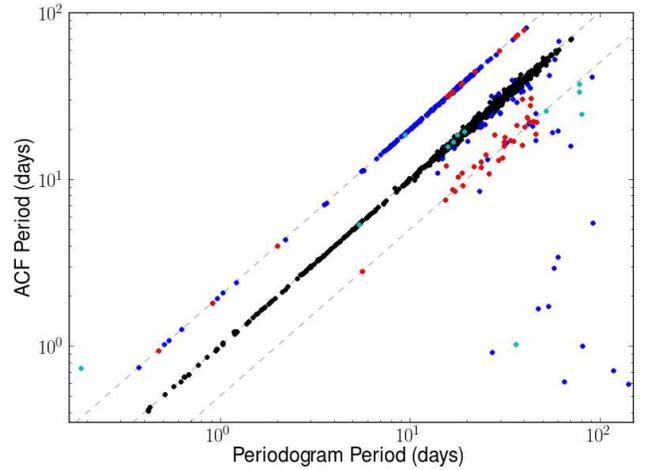}
  \caption{Comparison of the periods detected using the ACF and the
    periodogram, for all stars with a visually confirmed periodicity and 
    ACF detection. The correct ACF detections where the periodogram
    detections are within 10\% of the ACF period are shown in black (1298), the cases where 
    the periodogram period is not within 10\% of the ACF period are shown in dark blue (215). 
    Cases where the ACF initially selected the wrong period but the 
    periodogram selected the correct one are
    marked by red points (45). Cases where both the ACF and periodogram 
    detected the wrong period are marked as cyan points (12).
    The grey dashed lines show where $P_{\rm ACF} = P_{\rm LS}$, 
    $P_{\rm ACF} = 2\times P_{\rm LS}$ and $P_{\rm ACF} = 0.5 \times P_{\rm LS}$.}
  \label{fig:acf_sine}
\end{figure}

In 222 (14\%) of the periodic cases, the periodogram
does not detect the correct period (to within 10\%). The most common discrepancy arises 
from cases where the periodogram has detected half the period of the 
ACF (170 cases). We stress that there is no obvious way of identifying
these cases automatically using the periodogram alone (for example, we
tried to use the relative heights of the periodogram peaks, without success).
In 11 cases the periodogram selects erroneous long period peaks and the ACF
selects the correct shorter period peak.

On the other hand, the ACF yielded 57 incorrect ACF period detections (3.6\%), 45 cases occur 
where the ACF method selects the wrong period, and the periodogram
selects the correct period. These cases are marked as red points in the 
Figure~\ref{fig:acf_sine}. For the remaining 12, both the ACF and 
periodogram detect the wrong period (cyan points). 

\begin{figure*}
  \centering
  \includegraphics[width=0.55\linewidth]{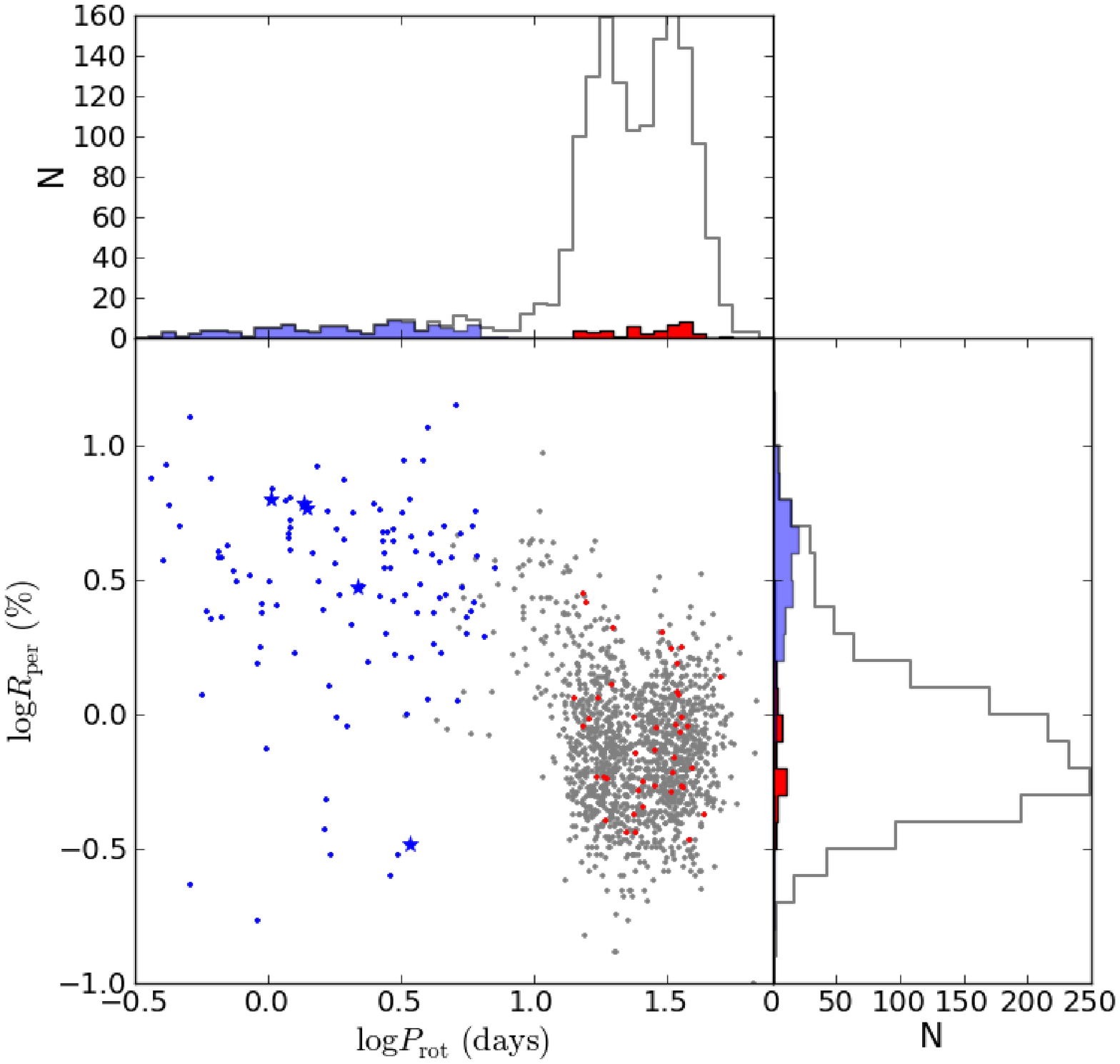} 
  \caption{Period versus amplitude for the rotating \protect\kepler\
    field M-dwarfs. Blue dots represent objects with $\prot<10$\,days,
    which also display unusually stable modulation patterns in their
    light curves, and blue stars known, short-period eclipsing
    binaries \protect\citep{prs+11}. Red dots represent the host stars
    of candidate transiting planets \protect\citep{bat+12}. All the
    other M-dwarfs with detected rotation periods are shown as grey
    dots. The histograms of each parameter are shown along the
    corresponding axis, with matching colours. Two long-period binaries
    are not shown as blue stars in the plot.}
  \label{fig:per_amp}
\end{figure*}

\begin{figure}
  \centering
  \includegraphics[width=\linewidth]{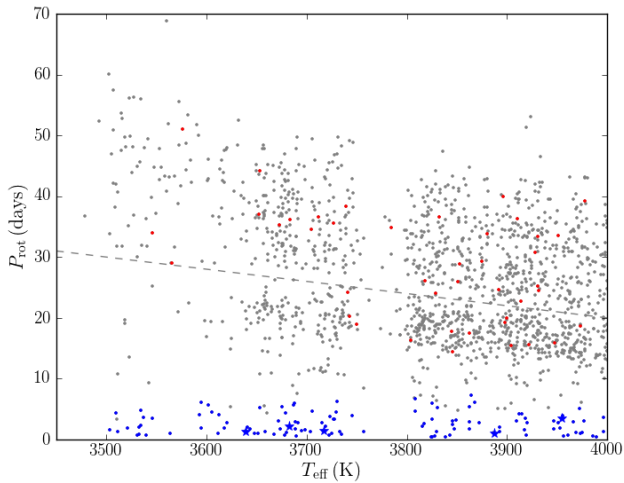} 
  \caption{Period versus effective temperature for the rotating
    \protect\kepler\ field M-dwarfs. The symbols and colours as the
    same as in
    Fig.~\protect\ref{fig:per_amp}. \protect\ The
      apparent dearth of objects around $T_{\rm eff}=3775$\,K is not
      apparent in any of the KIC colours, and is therefore thought to
      be a result of the KIC stellar parameter estimation procedure,
      rather than a real effect. The dashed line marks the 
      location of the cut made by eye between the fast and slow rotators.}
  \label{fig:per_teff}
\end{figure}

In one case (KIC\,10553513, not included in the periodic sample), the
large residual systematics prevent either the ACF or the periodogram 
from detecting the rotation period which is visible in the light curve.
The only case for which the periodogram is able to detect the period
and the ACF is not (even with manual correction) is KIC\,5480340 
(not included in the periodic sample). This is due to the extremely 
short period (0.25 days). The steep gradient in this region of
the ACF prevents peak detection, which we plan to resolve in a
future version of the algorithm.

We conclude that the periodogram remains a valuable technique
for period detection, however the clarity and robustness of the ACF 
method makes it our tool of choice for the measurement of stellar rotation 
in light curves. The ACF method works independently of periodogram based
methods and is ideal for determining rotation period statistics for large datasets.

\section{Discussion}
\label{sec:disc}
\subsection{Period, amplitude and temperature}

We now examine the period distribution of our sample as a function of
amplitude (Fig.~\ref{fig:per_amp}) and effective temperature
(Fig.~\ref{fig:per_teff}). A number of interesting features are
immediately apparent. 

First, the period distribution is clearly bimodal for
$\prot>10$\,days, with peaks of approximately equal height at $\sim
19$ and $\sim 33$\,days. This bimodality appears statistically
significant in log period space, with a Hartigan's dip test $p$-value of
0.0003 \citep{har85}. The two peaks of the
rotation period distribution form two distinct sequences in
period-temperature space (Fig.~\ref{fig:per_teff}). The period
decreases with increasing temperature, in much the same way, for both
sequences. This variation as a function of temperature
implies the gap between the sequences is not a result of
systematics in the \kepler\ light curves leading to missed 
detections in a particular period range. We define the short and long period samples 
based on the line, plotted somewhat arbitrarily, shown in Fig.~\ref{fig:per_teff}.

\begin{figure}
  \centering
  \includegraphics[width=\linewidth]{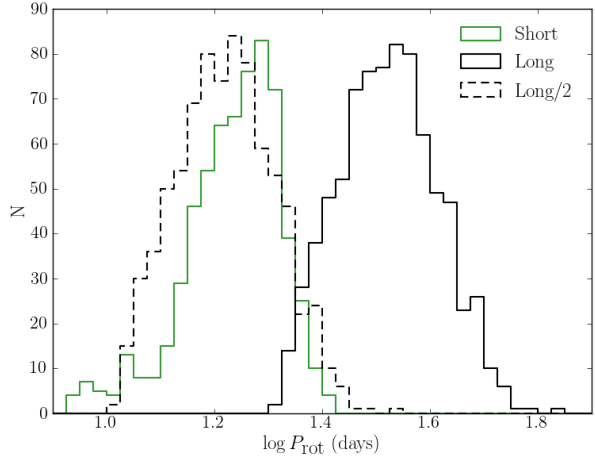} 
  \caption{Histogram of the short and long period M-dwarfs, 
  as defined by the cut marked with the dashed line in Fig.~\ref{fig:per_teff}. 
  The dashed histogram marks the long periods divided by 2, and provides
  further evidence that the short period set are unlikely to be erroneous
  half-period measurements of the long period population. }
  \label{fig:rot_hist}
\end{figure}

\begin{figure}
  \centering
  \includegraphics[width=\linewidth]{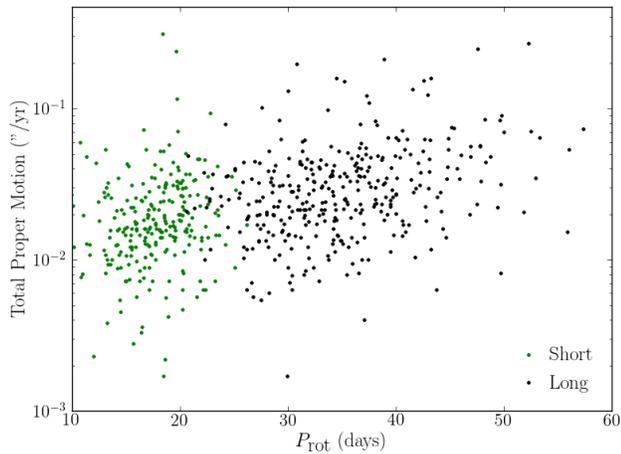} 
  \caption{Total proper motion against Period for the
  short period (green) and long period (black) M-dwarfs, as defined by the cut marked
  with the dashed line in Fig.~\ref{fig:per_teff}. }
  \label{fig:pm_scatter}
\end{figure}

\begin{figure}
  \centering
  \includegraphics[width=\linewidth]{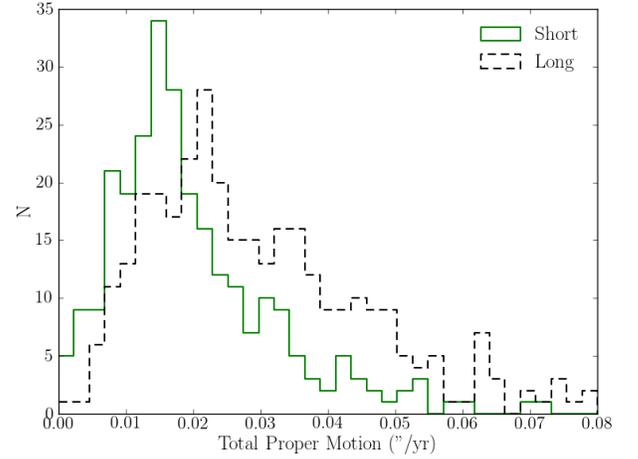} 
  \caption{Histogram of the total proper motion for the
  short period and long period M-dwarfs, as defined by the cut marked
  with the dashed line in Fig.~\ref{fig:per_teff}. The uncertainty on the 
  proper motion values is 0.02"/year but given the high numbers in 
  each sample these results are still significant. The known and 
  `possible' binaries have been removed from this sample (blue
  points and stars in Fig.~\ref{fig:per_teff}).}
  \label{fig:pm_hist}
\end{figure}
 
Before interpreting this result, we must address the possibility that
the bimodal distribution is spurious: an error of a factor of two in
the periods of about half the objects could give rise to the observed
distribution. Such errors are not uncommon in rotational studies based
on ground-based data \citep[see e.g.][]{col+09}. However, they are
less likely to be prevalent in the present study, for the following
reasons. We can exclude a scenario where we measured twice the true
period for the objects belonging to the longer period peak, because of
the continuous sampling of our data. The alternative scenario is that
we measured half the true period for the objects belonging to the
shorter period peak. This kind of problem arises when the brightness
distribution of the stellar surface is bimodal, due to concentrations
of active regions on opposite hemispheres, resulting in `double-dip'
light curves. However, our ACF peak selection routine was specifically
designed to address this problem -- this is one of the strengths of
the method. In fact, for 210 of the objects belonging to the shorter
period peak, a `double-dip' light curve lead to the second ACF peak 
being automatically selected as the rotation period. 
Unless these stars have quadrupolar surface brightnesses,
it is very unlikely that we underestimated their periods by a factor
of two.

The case for genuine bimodality is further strengthened by comparing
the detected distribution to that which would arise if either set were 
wrong by a factor of 2. Fig.~\ref{fig:rot_hist} shows that division of the
long period set by 2 does not reproduce the short period set (and equivalently 
neither does multiplication of the short period group). 
A quantitative two-sample Kolmogorov-Smirnov (KS) test of the short period
sample and the long period sample divided by 2 gives a p-value
$9\times10^{-14}$, confirming they are very unlikely to be
drawn from the same distribution.

Fig.~\ref{fig:pm_scatter} shows a weak correlation between
period and total proper motion for the two samples. 
We also see tentative evidence of a difference in the proper
motion distributions of the short and long period samples. Fig.~\ref{fig:pm_hist}
shows the histograms for stars in the short and long period samples
with non-zero proper motion KIC measurements (266 short, 345 long). The proper motion 
values from the KIC are taken from a selection of catalogs{\footnote{\kepler\ 
Stellar Classification Program, Hipparcos, Tycho-2, UCAC2, 2MASS and 
USNO-B1.0.} where available. Total proper motion is listed on NStED as having accuracy of 
20 milliarcseconds per year, but the large number of objects in each sample support 
the difference between them.} 

We also checked for differences in the galactic latitude and \kepler\ magnitude
distributions of the two samples but did not see any clear differences. By using the
galactic latitude measurements and an approximate distance estimate
from average apparent and observed M-dwarf magnitudes, we conclude that
the observed population lies within a small fraction of the scale height of the 
thin disk, and therefore a variation of properties with Galactic latitude is not expected.

For the remainder of this study, we therefore assume our sample 
contains two distinct stellar populations. The gap between the two sequences in 
Fig.~\ref{fig:per_teff} is reminiscent of the Vaughan-Preston gap (hereafter, V-P gap) found by 
\cite[][hereafter, VP80]{vau+80} when studying the chromospheric activity levels of a relatively small sample of F and G stars as a function of $B-V$ colour. As chromospheric activity, like stellar rotation, declines with age, both observed gaps suggest two waves of star formation in the Solar neighbourhood, as discussed by VP80 themselves and by numerous authors subsequently (see, for example, 
\citealt{bar88}, \citealt{pac+04} and references therein). Alternative explanations for the V-P gap include a discontinuity in the chromospheric activity level or of the rotation period at a particular point in the star's evolution. \cite{noy+84} found no evidence for a difference in the relationship between chromospheric activity, rotation period and colour for stars on either side of the V-P gap, which led them to dismiss the first of these alternatives. Our results also suggest the gap does not result from a discontinuity in the chromospheric activity level, as we see no obvious difference in the amplitude of photometric variability of the slow and fast rotators. Of the two remaining hypotheses, that of two waves of star formation seems more plausible, and is supported by the tentative differences we observe between the proper motion distributions of the two groups, but more kinematic data would be needed to confirm different epochs of star formation.

The fact that these have  different median periods \emph{and} different proper motion 
characteristics suggests that they may result from two distinct waves of 
star formation: as already discussed, low-mass main-sequence stars 
spin down as they age, and older populations tend to have larger 
velocity dispersions than younger populations, due to dynamical 
heating of the Galactic disk over long timescales \citep[see][and references therein]{fre+02}. 
The ratio of the two peaks in the period distribution
corresponds to an age ratio of $\sim 3$, if one assumes that M-dwarfs
spin down as $t^{0.5}$, as observed for Sun-like stars
\citep{sku72}. Recent rotational studies of low-mass stars in
open-clusters suggest that the index of the spin-down low might be
closer to $0.6$ \citep[see e.g.][]{mei+09}, which would imply an age
ratio of $\sim 2.5$. We note that an interpretation in terms of thin
and thick disk populations is unlikely because there should be many
fewer thick disk than thick objects in the \kepler\ sample, which is
essentially magnitude-limited. 

Further kinematic and distance information is required to draw stronger conclusions
about the nature of the two populations. 
The \kepler\ mission itself may, in the medium term, provide this
information: the pixel-level data can be
used to monitor the centroid of each star over the full lifetime of
the mission, which should enable the measurement of proper motions and
parallaxes for some targets. \cite{mon+10} shows that astrometric precision 
for a single 30 minute measure is $\textless\ 4$ milliarcseconds. 
In the longer term, the GAIA mission \citep{deb+12} will
provide this information with greater precision and for more objects.

While the bulk of the objects (89\%) have periods in the range 10--50\,days,
a small fraction ($8\%$) of the periods are $<7$\,days. Almost all
these rapid rotators display unusually stable light curves over the
full 10 month dataset (blue points in Figs.~\ref{fig:per_amp} and
\ref{fig:per_teff}, label `PB' in Table~\ref{tab:prot}). Most of them also have relatively large amplitudes
($>2\%$, compared to 0.2--2\% for the bulk of the sample). Two
possible explanations spring to mind for these rapid rotators: they
may be significantly younger than the rest of the sample (for example
members of the young disk population, see
Section~\ref{sec:period_mass}), or they may be close binary systems,
which have become rotators due to spin-orbit interactions. 
The 5 known short-period eclipsing binaries in our sample 
(shown as stars in Fig.~\ref{fig:per_amp}) all display synchronised 
rotation signals, suggesting that  at least some of the other 
However, we cannot conclusively distinguish between
the two possibilities without spectroscopy.

\begin{figure*}
  \centering
  \includegraphics[width=0.7\linewidth]{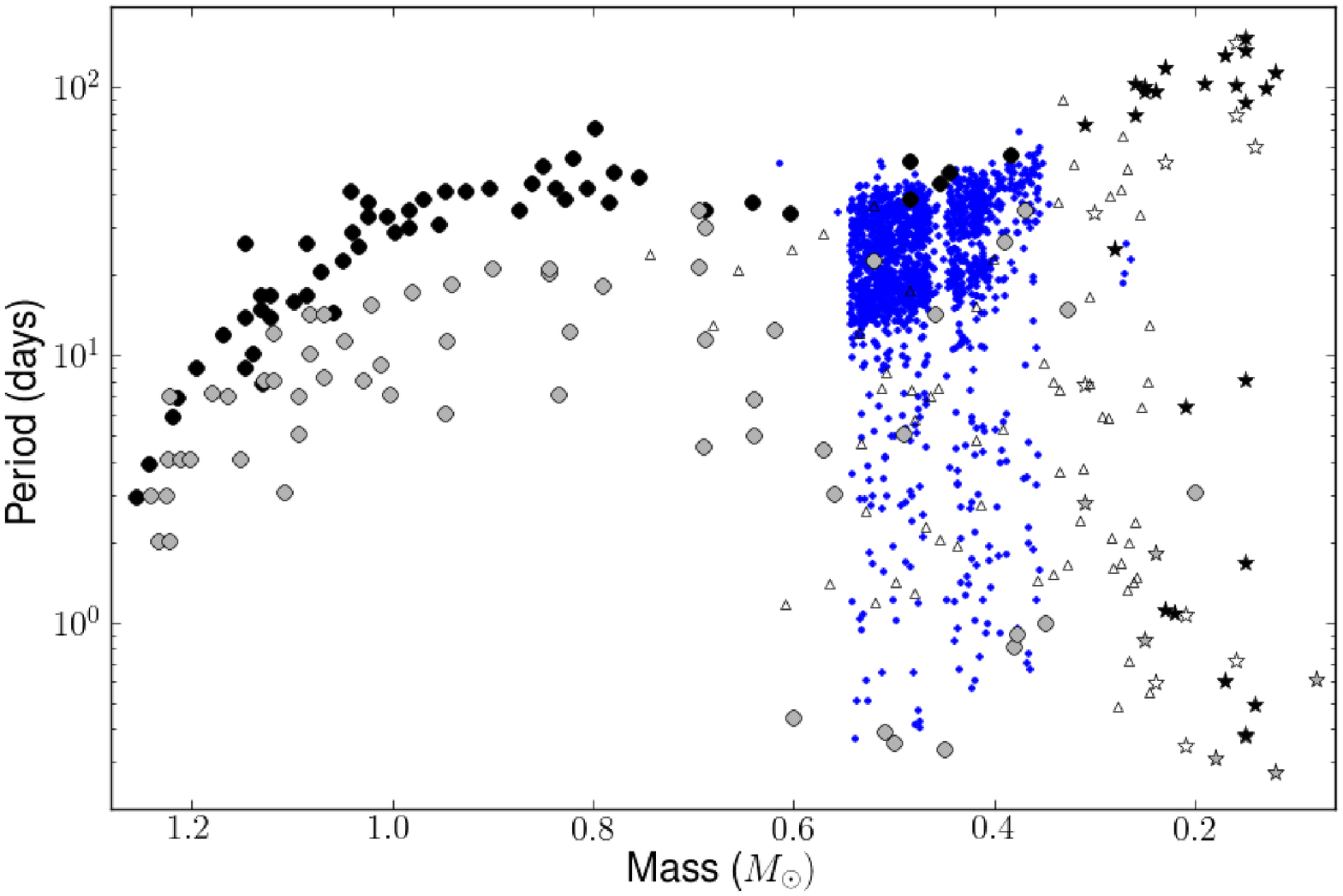} 
  \caption{Period versus mass for low-mass field stars. Based on data
    from \citet{bal+96} and \citet{kist07} (circles) and \mearth\ data from
    \citet{irw+11} (stars), with grey and black symbols representing
    objects with young and old disk kinematics,
    respectively. Additional M-dwarf periods
    from the WFCAM Transit Survey \citep{gou+12}, for which no
    kinematic classification is available, are shown as white
    triangles. The new results from the present study are shown as blue points.}
  \label{fig:per_mass}
\end{figure*}

Finally, all the planet-host candidates (\citealt{bat+12}, red points
in Figs.~\ref{fig:per_amp} and \ref{fig:per_teff}) have $\prot >
10$\,days and amplitude $<2\%$. If the period and amplitude
distributions of the planet-host candidates were the same as those of
the rest of the sample, we would have expected $\sim 3$ of them to
have $\prot < 10$\,days and/or amplitude $>2\%$ (of all the M-dwarfs
with rotation periods, $10\%$ fall outside these limits, and there are
42 candidate planet host stars in the sample). The number of objects
concerned is too small to draw firm conclusions, but it does suggest
that either the search for transits in \kepler\ light curves may be
less complete around active, rapidly rotating stars, or that there are
fewer transiting planets around these stars (which could be the case
of the rapid rotators are close binaries). Either way, this should be
taken into account when inferring the incidence of planets around
different types of stars.

\subsection{Period-mass relation}
\label{sec:period_mass}

To compare our results with other rotational studies of low-mass
stars, we estimated masses for the \kepler\ M-dwarfs from the KIC
effective temperatures. Much of the previous work on rotational
evolution \citep{kaw89,bar03, bar07,mei+09,col+09} uses a directly
observed colour index, such as $B-V$ or $J-K$, instead of
model-dependent mass estimates, to confront models with observations.
We opted for masses for three reasons. First, angular momentum
evolution models generally depend on fundamental parameters rather
than colours. Second, neither of the colour indices most frequently
used, $B-V$ and $J-K$, is well matched to our targets, which are very
faint in $B$, and have almost constant $J-K$ over the mass range of
interest (in contrast to G and K-dwarfs, where $J-K$ is a steeper
function of mass). Third, stellar parameters based on multi-colour
photometry should be less sensitive to reddening than any single colour
index.

The masses used in this study were obtained by interpolating the
$630$\,Myr isochrone of \cite{bar+98}. We checked that the results are
essentially unchanged if the age is increased by a factor of up to
10. While useful for a general comparison, we stress that these masses
are very uncertain and should not be used for individual objects. The
$200\,K$ formal uncertainties on $T_{\rm eff}$ alone translate into
mass uncertainties ranging from $0.15\,M_\odot$ (for $M \sim
0.3\,M_\odot$) to $0.1\,M_\odot$ (for $M \sim
0.6\,M_\odot$). Additionally, any systematic errors in $T_{\rm eff}$,
discussed in Section~\ref{sec:tar_sel} are automatically
propagated through to the masses. 

Fig.~\ref{fig:per_mass} shows the period-mass relation for field
stars with masses below $1.3\,M_\odot$, based on a compilation of
literature sources \citep{bal+96, kist07, irw+11, gou+12}. Our sample (small blue dots on
Fig.~\ref{fig:per_mass}) ties in smoothly with the existing data at
both higher and lower masses (circles, stars and triangles).
The tight sequence followed by old F, G and K stars is clearly seen on
Fig.~\ref{fig:per_mass}, as are the younger stars gradually
evolving towards that sequence. This evolution is relatively well
reproduced by simple models of angular momentum loss via a magnetised
wind \citep{kaw88,kaw89}, and forms the basis for rotational
age-dating, or gyrochronology \citep{bar03,bar07}.

The picture is more complex for M-dwarfs, which evolve more slowly,
and hence have not converged onto a common sequence, even after
several Gyr. However, the new, much larger sample of early M-dwarfs
strengthens two interesting features, which were only hinted at by
previously published results.

First, we might have expected to see some kind of transition around
$0.35\,M_\odot$ \citep{cha+97, sch+11}, which corresponds to the transition between fully
convective stars and stars with a radiative core, because the
so-called interface dynamo, which powers the magnetic field of Sun-like
stars, cannot operate in the absence of a radiative core. No such transition
is seen among the slowly rotating stars, which define the upper
envelope of the period-mass relation. There does seem to be a transition at $\sim
0.35\,M_\odot$ for rapid rotators: below this mass, rapid rotators
persist even among kinematically old stars, as the \mearth\ sample of
\citet{irw+11} illustrates, but above it, all the fast rotators are
kinematically young, as \cite{kist07}
pointed out. The \kepler\ sample as it stands cannot shed much
additional light on this point because we currently lack kinematic and
distance information.

Second, a transition that was not expected, but is clearly seen,
occurs in the upper envelope of the period-mass relation around
$0.6$--$0.55\,M_\odot$, where the slope of the relation suddenly
changes sign. To our knowledge, this intriguing feature has not yet been
discussed in the literature, and it will need to be accounted for in future
modelling work. In particular, this transition will need to be
incorporated into empirical gyrochronological relations, if the latter are to be
applied to M-dwarfs. 

The moderately slow rotation of field M-dwarfs has been problematic
for some time. The latest generation of theoretical models of angular
momentum loss via a magnetised wind \citep{rei+12} naturally explains
the wide range of rotation rates observed for M-dwarfs in early
main-sequence clusters. However, the behaviour of field M-dwarfs can
only be reproduced by assuming that the critical rotation rate (above
which the magnetic field saturates) is not universal, but depends on
mass and perhaps even on age. With the new results, upper-envelope of
the period-mass relation is now much better defined, which should
prove valuable in testing and calibrating future refinements of the
models.

\section{Conclusions}
\label{sec:conclusions}

We report rotation periods for 1570 main-sequence M-dwarf 
stars with masses between 0.3 and 0.55\,$M_\odot$,
measured using a new method based on the autocorrelation function,
applied to the first 10 months of data from \kepler. 
The fraction of objects in which we detected periods, 63.2\%, is
remarkably high. For comparison, \cite{irw+11} detected periods in
15\% of their 273-object sample. The contrast can be explained by the
unprecedented precision and continuous sampling of the \kepler\ data,
and by the performance of the ACF method. 

The bulk of our period detections fall into two distinct groups, with
periods in the range 10--25 and 25--80 days with peaks at $\sim 19$ and $\sim
33$\,days respectively. We suggest that these correspond to two 
stellar populations with different median ages. The comparison 
of available non-zero proper motions for the 
two main samples further supports the 
hypothesis that they have different ages. The more slowly 
rotating group has typically higher proper motion values,
as would be expected for an older population.
Within each group there is also a
weak correlation between period and proper motion. This study shows
the potential use of variability statistics as a probe of star formation
history, although further data on distance and kinematics are
required to draw sound conclusions about the nature of these
regions.

In each of the two groups of stars, which form the two peaks of the period distribution, 
the rotation period tends to increase with decreasing mass. 
This trend is very clear for the upper envelope of the relation, 
but we also checked that it applies to the bulk of the stars by examining 
the median of the rotation periods for each group (slow and fast rotators) 
in $0.05\,M_\odot$ bins. By combining our 
results existing rotation data from the literature 
over the mass range $0.1$--$1.3\,M_\odot$, we have 
shown that this relation extends over the whole M-dwarf regime 
($0.1$--$0.55\,M_\odot$) but is in stark contrast to the behaviour 
of K-stars, whose period decreases with decreasing mass. 
To our knowledge, this dichotomy between K and M-dwarfs 
has not been noticed before, and is not predicted by current models.

A small fraction ($\sim8$\%) of the M-dwarfs display short
($< 7$\,days), stable periods, and marginally enhanced variability
compared to the rest of the sample. The most likely explanations are
that these are short-period binaries or very young stars.

When combining the \kepler\ sample with data from the literature, we see no 
evidence for a break in the period-mass relation around
$0.35\,M_\odot$, even though stars below this mass are expected to
remain fully convective. Such a break might have been expected if the
development of a radiative core played a key role in driving a
large-scale magnetic field, as proposed for example by
\citet{bar03}. We note that the \kepler\ and \mearth\ field M-dwarfs
lie in the `rotational gap' defined by \citet{bar10}. This suggests
that the Rossby number, which plays a key role in controlling the
rotational evolution of G and K stars, may be less important for low
mass stars. 

Analysis of additional quarters of \kepler\ data may reveal 
long periods for some of the objects without detections 
in the present sample. We are also working to improve the 
systematics correction with an alternative to the PDC-MAP
(Roberts et al. submitted) , and to enhance the performance
of the ACF method. Residual systematics and quarter joins are the limiting
factors in the current analysis because they introduce steep variations 
in the ACF. By optimising the ACF smoothing parameters after initial 
period detection, and working on methods to separate the periodic signal from the 
long term ACF trends, we hope reach a higher level of precision and
clarity.

This study is the first large-scale investigation of rotation in field
M-dwarfs, and provides the first useful constraints on the period-mass
relation for these objects after they have settled onto a common
rotational sequence.  This work also demonstrates the power of
\kepler\ data and of the ACF method for rotation studies, and paves
the way for a truly systematic survey of rotation rates on the main
sequence, from mid-F to mid-M spectral types, which we plan to examine
in future papers. We will also seek to automate the ACF period
verification stages for use on larger samples of targets, and run
extensive simulations in order to quantify detection efficiency.

\section*{Acknowledgments}
The authors wish to express their special thanks to the \kepler\
Science Operations Centre and pipeline teams, whose dedicated efforts
have produced an extremely valuable resource for the stellar
astrophysics community.  We are also grateful to All Souls College for
electing TM to a visiting fellowship, without which this project would
not have been initiated.

This paper includes data collected by the \kepler\ mission. Funding
for the \kepler\ mission is provided by the NASA Science Mission
Directorate. All of the data presented in this paper were obtained
from the Mikulski Archive for Space Telescopes (MAST). STScI is
operated by the Association of Universities for Research in Astronomy,
Inc., under NASA contract NAS5-26555. Support for MAST for non-HST
data is provided by the NASA Office of Space Science via grant
NNX09AF08G and by other grants and contracts. AM and SA acknowledge
support from UK Science and Technology Facilities Council (grant refs
ST/F006888/1 and ST/G002266/2). The research leading to these results
has received funding from the European Research Council under the EU's
Seventh Framework Programme (FP7/(2007-2013)/ ERC Grant Agreement
No. 291352).

\bibliography{mrot}

\begin{thebibliography}{54}
\expandafter\ifx\csname natexlab\endcsname\relax\def\natexlab#1{#1}\fi

\bibitem[{Affer {et~al}\mbox{.}(2012)Affer, Micela, Favata, \&
  Flaccomio}]{aff+12}
Affer L., Micela G., Favata F., Flaccomio E., 2012, MNRAS, 424, 11

\bibitem[{{Aigrain} {et~al}\mbox{.}(2012){Aigrain}, {Pont}, \&
  {Zucker}}]{aig+12}
{Aigrain} S., {Pont} F., {Zucker} S., 2012, MNRAS, 419, 3147

\bibitem[{{Baliunas} {et~al}\mbox{.}(1996){Baliunas}, {Sokoloff}, \&
  {Soon}}]{bal+96}
{Baliunas} S., {Sokoloff} D., {Soon} W., 1996, ApJL, 457, L99

\bibitem[{{Baraffe} {et~al}\mbox{.}(1998){Baraffe}, {Chabrier}, {Allard}, \&
  {Hauschildt}}]{bar+98}
{Baraffe} I., {Chabrier} G., {Allard} F., {Hauschildt} P.~H., 1998, A\&A, 337,
  403

\bibitem[{Barnes(2003)}]{bar03}
Barnes S.~A., 2003, ApJ, 586, 464

\bibitem[{Barnes(2007)}]{bar07}
Barnes S.~A., 2007, ApJ, 669, 1167

\bibitem[{Barnes(2010)}]{bar10}
Barnes S.~A., 2010, ApJ, 722, 222

\bibitem[{Barnes \& Kim(2010)}]{bar+10}
Barnes S.~A., Kim Y.-C., 2010, ApJ, 721, 675

\bibitem[{{Barry}(1988)}]{bar88}
{Barry} D.~C., 1988, ApJ, 334, 436

\bibitem[{{Basri} {et~al}\mbox{.}(2011){Basri}, {Walkowicz}, {Batalha},
  {Gilliland}, {Jenkins}, {Borucki}, {Koch}, {Caldwell}, {Dupree}, {Latham},
  {Marcy}, {Meibom}, \& {Brown}}]{bas+11}
{Basri} G. {et~al.}, 2011, AJ, 141, 20

\bibitem[{{Basri} {et~al}\mbox{.}(2010){Basri}, {Walkowicz}, {Batalha},
  {Gilliland}, {Jenkins}, {Borucki}, {Koch}, {Caldwell}, {Dupree}, {Latham},
  {Meibom}, {Howell}, \& {Brown}}]{bas+10}
{Basri} G. {et~al.}, 2010, ApJL, 713, L155

\bibitem[{{Batalha} {et~al}\mbox{.}(2010){Batalha}, {Borucki}, {Koch},
  {Bryson}, {Haas}, {Brown}, {Caldwell}, {Hall}, {Gilliland}, {Latham},
  {Meibom}, \& {Monet}}]{bat+10}
{Batalha} N.~M. {et~al.}, 2010, ApJL, 713, L109

\bibitem[{{Batalha} {et~al}\mbox{.}(2012){Batalha}, {Rowe}, {Bryson},
  {Barclay}, {Burke}, {Caldwell}, {Christiansen}, {Mullally}, {Thompson},
  {Brown}, {Dupree}, {Fabrycky}, {Ford}, {Fortney}, {Gilliland}, {Isaacson},
  {Latham}, {Marcy}, {Quinn}, {Ragozzine}, {Shporer}, {Borucki}, {Ciardi},
  {Gautier}, {Haas}, {Jenkins}, {Koch}, {Lissauer}, {Rapin}, {Basri}, {Boss},
  {Buchhave}, {Charbonneau}, {Christensen-Dalsgaard}, {Clarke}, {Cochran},
  {Demory}, {Devore}, {Esquerdo}, {Everett}, {Fressin}, {Geary}, {Girouard},
  {Gould}, {Hall}, {Holman}, {Howard}, {Howell}, {Ibrahim}, {Kinemuchi},
  {Kjeldsen}, {Klaus}, {Li}, {Lucas}, {Morris}, {Prsa}, {Quintana},
  {Sanderfer}, {Sasselov}, {Seader}, {Smith}, {Steffen}, {Still}, {Stumpe},
  {Tarter}, {Tenenbaum}, {Torres}, {Twicken}, {Uddin}, {Van Cleve},
  {Walkowicz}, \& {Welsh}}]{bat+12}
{Batalha} N.~M. {et~al.}, 2012, ApJ, submitted, available at arXiv:1202.5852

\bibitem[{Borucki {et~al}\mbox{.}(2010)Borucki, Koch, Basri, Batalha, Brown,
  Caldwell, Caldwell, Christensen-Dalsgaard, Cochran, DeVore, Dunham, Dupree,
  Gautier, Geary, Gilliland, Gould, Howell, Jenkins, Kondo, Latham, Marcy,
  Meibom, Kjeldsen, Lissauer, Monet, Morrison, Sasselov, Tarter, Boss,
  Brownlee, Owen, Buzasi, Charbonneau, Doyle, Fortney, Ford, Holman, Seager,
  Steffen, Welsh, Rowe, Anderson, Buchhave, Ciardi, Walkowicz, Sherry, Horch,
  Isaacson, Everett, Fischer, Torres, Johnson, Endl, MacQueen, Bryson, Dotson,
  Haas, Kolodziejczak, Van~Cleve, Chandrasekaran, Twicken, Quintana, Clarke,
  Allen, Li, Wu, Tenenbaum, Verner, Bruhweiler, Barnes, \& Prsa}]{bor+10}
Borucki W.~J. {et~al.}, 2010, Science, 327, 977

\bibitem[{{Borucki} {et~al}\mbox{.}(2011){Borucki}, {Koch}, {Basri}, {Batalha},
  {Brown}, {Bryson}, {Caldwell}, {Christensen-Dalsgaard}, {Cochran}, {DeVore},
  {Dunham}, {Gautier}, {Geary}, {Gilliland}, {Gould}, {Howell}, {Jenkins},
  {Latham}, {Lissauer}, {Marcy}, {Rowe}, {Sasselov}, {Boss}, {Charbonneau},
  {Ciardi}, {Doyle}, {Dupree}, {Ford}, {Fortney}, {Holman}, {Seager},
  {Steffen}, {Tarter}, {Welsh}, {Allen}, {Buchhave}, {Christiansen}, {Clarke},
  {Das}, {D{\'e}sert}, {Endl}, {Fabrycky}, {Fressin}, {Haas}, {Horch},
  {Howard}, {Isaacson}, {Kjeldsen}, {Kolodziejczak}, {Kulesa}, {Li}, {Lucas},
  {Machalek}, {McCarthy}, {MacQueen}, {Meibom}, {Miquel}, {Prsa}, {Quinn},
  {Quintana}, {Ragozzine}, {Sherry}, {Shporer}, {Tenenbaum}, {Torres},
  {Twicken}, {Van Cleve}, {Walkowicz}, {Witteborn}, \& {Still}}]{bor+11}
{Borucki} W.~J. {et~al.}, 2011, ApJ, 736, 19

\bibitem[{Bouvier {et~al}\mbox{.}(1997)Bouvier, Forestini, \& Allain}]{bou+97}
Bouvier J., Forestini M., Allain S., 1997, A\&A, 326, 1023

\bibitem[{Brown {et~al}\mbox{.}(2011)Brown, Latham, Everett, \&
  Esquerdo}]{bro+11}
Brown T.~M., Latham D.~W., Everett M.~E., Esquerdo G.~A., 2011, AJ, 142, 112

\bibitem[{{Castelli} \& {Kurucz}(2004)}]{ck04}
{Castelli} F., {Kurucz} R.~L., 2004, in IAU Symp., Vol. 210, Modelling of
  Stellar Atmospheres, available at arXiv:astro-ph/0405087

\bibitem[{{Chabrier} \& {Baraffe}(1997)}]{cha+97}
{Chabrier} G., {Baraffe} I., 1997, A\&A, 327, 1039

\bibitem[{Ciardi {et~al}\mbox{.}(2011)Ciardi, von Braun, Bryden, van Eyken,
  Howell, Kane, Plavchan, Ram{\'\i}rez, \& Stauffer}]{cia+11}
Ciardi D.~R. {et~al.}, 2011, AJ, 141, 108

\bibitem[{Collier~Cameron {et~al}\mbox{.}(2009)Collier~Cameron, Davidson, Hebb,
  Skinner, Anderson, Christian, Clarkson, Enoch, Irwin, Joshi, Haswell,
  Hellier, Horne, Kane, Lister, Maxted, Norton, Parley, Pollacco, Ryans,
  Scholz, Skillen, Smalley, Street, West, Wilson, \& Wheatley}]{col+09}
Collier~Cameron A. {et~al.}, 2009, MNRAS, 400, 451

\bibitem[{{de Bruijne}(2012)}]{deb+12}
{de Bruijne} J.~H.~J., 2012, Ap\&SS, 341, 31

\bibitem[{{Freeman} \& {Bland-Hawthorn}(2002)}]{fre+02}
{Freeman} K., {Bland-Hawthorn} J., 2002, ARA\&A, 40, 487

\bibitem[{{Goulding} {et~al}\mbox{.}(2012){Goulding}, {Barnes}, {Pinfield},
  {Kov{\'a}cs}, {Birkby}, {Hodgkin}, {Catal{\'a}n}, {Sip{\H o}cz}, {Jones},
  {Del Burgo}, {Jeffers}, {Nefs}, {G{\'a}lvez-Ortiz}, \& {Martin}}]{gou+12}
{Goulding} N.~T. {et~al.}, 2012, MNRAS, 427, 3358

\bibitem[{{Hall} \& {Henry}(1994)}]{hahe94}
{Hall} D.~S., {Henry} G.~W., 1994, International Amateur-Professional
  Photoelectric Photometry Communications, 55, 51

\bibitem[{{Harrison} {et~al}\mbox{.}(2012){Harrison}, {Coughlin}, {Ule}, \&
  {L{\'o}pez-Morales}}]{har+12}
{Harrison} T.~E., {Coughlin} J.~L., {Ule} N.~M., {L{\'o}pez-Morales} M., 2012,
  AJ, 143, 4

\bibitem[{{Hartigan} \& {Hartigan}(1985)}]{har85}
{Hartigan} J.~A., {Hartigan} P.~M., 1985, The Annals of Statistics, 13, 70

\bibitem[{{Irwin} {et~al}\mbox{.}(2006){Irwin}, {Aigrain}, {Hodgkin}, {Irwin},
  {Bouvier}, {Clarke}, {Hebb}, \& {Moraux}}]{irw+06}
{Irwin} J., {Aigrain} S., {Hodgkin} S., {Irwin} M., {Bouvier} J., {Clarke} C.,
  {Hebb} L., {Moraux} E., 2006, MNRAS, 370, 954

\bibitem[{Irwin {et~al}\mbox{.}(2011)Irwin, Berta, Burke, Charbonneau, Nutzman,
  West, \& Falco}]{irw+11}
Irwin J., Berta Z.~K., Burke C.~J., Charbonneau D., Nutzman P., West A.~A.,
  Falco E.~E., 2011, ApJ, 727, 56

\bibitem[{{Irwin} \& {Bouvier}(2009)}]{irw+09}
{Irwin} J., {Bouvier} J., 2009, in IAU Symp., Vol. 258, The Ages of Stars, pp.
  363--374

\bibitem[{{Jackson} \& {Jeffries}(2012)}]{jac+12}
{Jackson} R.~J., {Jeffries} R.~D., 2012, MNRAS, 423, 2966

\bibitem[{Kawaler(1988)}]{kaw88}
Kawaler S.~D., 1988, ApJ, 333, 236

\bibitem[{Kawaler(1989)}]{kaw89}
Kawaler S.~D., 1989, ApJ, 343, L65

\bibitem[{{Kiraga} \& {Stepien}(2007)}]{kist07}
{Kiraga} M., {Stepien} K., 2007, ACTAA, 57, 149

\bibitem[{{Mann} {et~al}\mbox{.}(2012){Mann}, {Gaidos}, {L{\'e}pine}, \&
  {Hilton}}]{man+12}
{Mann} A.~W., {Gaidos} E., {L{\'e}pine} S., {Hilton} E.~J., 2012, ApJ, 753, 90

\bibitem[{{McQuillan} {et~al}\mbox{.}(2012){McQuillan}, {Aigrain}, \&
  {Roberts}}]{mcq+11}
{McQuillan} A., {Aigrain} S., {Roberts} S., 2012, A\&A, 539, A137

\bibitem[{Meibom {et~al}\mbox{.}(2011)Meibom, Barnes, Latham, Batalha, Borucki,
  Koch, Basri, Walkowicz, Janes, Jenkins, Van~Cleve, Haas, Bryson, Dupree, F{\H
  u}r{\'e}sz, Szentgyorgyi, Buchhave, Clarke, Twicken, \& Quintana}]{mei+11}
Meibom S. {et~al.}, 2011, ApJL, 733, L9

\bibitem[{Meibom {et~al}\mbox{.}(2009)Meibom, Mathieu, \& Stassun}]{mei+09}
Meibom S., Mathieu R.~D., Stassun K.~G., 2009, ApJ, 695, 679

\bibitem[{{Monet} {et~al}\mbox{.}(2010){Monet}, {Jenkins}, {Dunham}, {Bryson},
  {Gilliland}, {Latham}, {Borucki}, \& {Koch}}]{mon+10}
{Monet} D.~G., {Jenkins} J.~M., {Dunham} E.~W., {Bryson} S.~T., {Gilliland}
  R.~L., {Latham} D.~W., {Borucki} W.~J., {Koch} D.~G., 2010, ApJ, submitted,
  available at arXiv:1203.1383

\bibitem[{{Mosser} {et~al}\mbox{.}(2009){Mosser}, {Baudin}, {Lanza}, {Hulot},
  {Catala}, {Baglin}, \& {Auvergne}}]{mos+09}
{Mosser} B., {Baudin} F., {Lanza} A.~F., {Hulot} J.~C., {Catala} C., {Baglin}
  A., {Auvergne} M., 2009, A\&A, 506, 245

\bibitem[{{Muirhead} {et~al}\mbox{.}(2012){Muirhead}, {Hamren}, {Schlawin},
  {Rojas-Ayala}, {Covey}, \& {Lloyd}}]{mui+12}
{Muirhead} P.~S., {Hamren} K., {Schlawin} E., {Rojas-Ayala} B., {Covey} K.~R.,
  {Lloyd} J.~P., 2012, ApJL, 750, L37

\bibitem[{{Noyes} {et~al}\mbox{.}(1984){Noyes}, {Hartmann}, {Baliunas},
  {Duncan}, \& {Vaughan}}]{noy+84}
{Noyes} R.~W., {Hartmann} L.~W., {Baliunas} S.~L., {Duncan} D.~K., {Vaughan}
  A.~H., 1984, ApJ, 279, 763

\bibitem[{{Pace} \& {Pasquini}(2004)}]{pac+04}
{Pace} G., {Pasquini} L., 2004, A\&A, 426, 1021

\bibitem[{{Pr{\v s}a} {et~al}\mbox{.}(2011){Pr{\v s}a}, {Batalha}, {Slawson},
  {Doyle}, {Welsh}, {Orosz}, {Seager}, {Rucker}, {Mjaseth}, {Engle}, {Conroy},
  {Jenkins}, {Caldwell}, {Koch}, \& {Borucki}}]{prs+11}
{Pr{\v s}a} A. {et~al.}, 2011, AJ, 141, 83

\bibitem[{{Reiners} \& {Mohanty}(2012)}]{rei+12}
{Reiners} A., {Mohanty} S., 2012, ApJ, 746, 43

\bibitem[{Scargle(1982)}]{sca82}
Scargle J.~D., 1982, ApJ, 263, 835

\bibitem[{{Scholz} {et~al}\mbox{.}(2011){Scholz}, {Irwin}, {Bouvier}, {Sip{\H
  o}cz}, {Hodgkin}, \& {Eisl{\"o}ffel}}]{sch+11}
{Scholz} A., {Irwin} J., {Bouvier} J., {Sip{\H o}cz} B.~M., {Hodgkin} S.,
  {Eisl{\"o}ffel} J., 2011, MNRAS, 413, 2595

\bibitem[{{Shumway} \& {Stoffer}(2010)}]{shst10}
{Shumway} R.~H., {Stoffer} D.~S., 2010, {Time Series Analysis and Its
  Applications: With R Examples}. Springer

\bibitem[{Skumanich(1972)}]{sku72}
Skumanich A., 1972, ApJ, 171, 565

\bibitem[{Smith {et~al}\mbox{.}(2012)Smith, Stumpe, van Cleve, Jenkins,
  Barclay, Fanelli, Girouard, Kolodziejczak, McCauliff, Morris, \&
  Twicken}]{smi+12}
Smith J.~C. {et~al.}, 2012, PASP, submitted, available at arXiv:1203.1383

\bibitem[{Stumpe {et~al}\mbox{.}(2012)Stumpe, Smith, van Cleve, Twicken,
  Barclay, Fanelli, Girouard, Jenkins, Kolodziejczak, McCauliff, \&
  Morris}]{stu+12}
Stumpe M.~C. {et~al.}, 2012, PASP, submitted, available at arXiv:1203.1382

\bibitem[{{Vaughan} \& {Preston}(1980)}]{vau+80}
{Vaughan} A.~H., {Preston} G.~W., 1980, PASP, 92, 385

\bibitem[{{Verner} {et~al}\mbox{.}(2011){Verner}, {Chaplin}, {Basu}, {Brown},
  {Hekker}, {Huber}, {Karoff}, {Mathur}, {Metcalfe}, {Mosser}, {Quirion},
  {Appourchaux}, {Bedding}, {Bruntt}, {Campante}, {Elsworth}, {Garc{\'{\i}}a},
  {Handberg}, {R{\'e}gulo}, {Roxburgh}, {Stello}, {Christensen-Dalsgaard},
  {Gilliland}, {Kawaler}, {Kjeldsen}, {Allen}, {Clarke}, \&
  {Girouard}}]{ver+11}
{Verner} G.~A. {et~al.}, 2011, APJL, 738, L28

\bibitem[{Zechmeister \& K{\"u}rster(2009)}]{zec+09}
Zechmeister M., K{\"u}rster M., 2009, A\&A, 496, 577

\end{thebibliography}
\bibliographystyle{mn2e_mod}

\label{lastpage}

\end{document}